\documentclass[headings,11pt]{llncs}

\usepackage[english]{babel}
\usepackage{graphicx}
\usepackage{amsmath}
\usepackage{fullpage}
\usepackage{mathptmx}
\usepackage{helvet} 
\usepackage[utf8]{inputenc}
\usepackage{color}
\usepackage{relsize}
\usepackage{amssymb}
\usepackage{array} 
\usepackage{rotating}
\usepackage{chapterbib}
\usepackage{lmodern}
\usepackage{xspace}
\usepackage{geometry}
\usepackage{yfonts}

\begin{document}

\title{Dynamic Provable Data Possession Protocols with Public Verifiability and Data Privacy}

\author{Cl\'{e}mentine Gritti$^1$, Rongmao Chen$^2$, Willy Susilo$^3$, Thomas Plantard$^3$}
\institute{$^1$EURECOM, Sophia Antipolis, France \\
\email{clementine.gritti@eurecom.fr} \\
$^2$College of Computer, National University of Defense Technology, China \\
\email{chromao@nudt.edu.cn} \\
$^3$Institute of Cybersecurity and Cryptology\\
School of Computing and Information Technology \\
University of Wollongong, Australia\\
\email{\{wsusilo,thomaspl\}@uow.edu.au}     
}

\maketitle

\begin{abstract}
Cloud storage services have become accessible and used by everyone.
Nevertheless, stored data are dependable on the behavior of the cloud servers, and losses and damages often occur.
One solution is to regularly audit the cloud servers in order to check the integrity of the stored data. 
The Dynamic Provable Data Possession scheme with Public Verifiability and Data Privacy presented in ACISP'15 is a straightforward design of such solution.
However, this scheme is threatened by several attacks.
In this paper, we carefully recall the definition of this scheme as well as explain how its security is dramatically menaced.
Moreover, we proposed two new constructions
for Dynamic Provable Data Possession scheme with Public Verifiability and Data Privacy based on the scheme presented in ACISP'15,
one using Index Hash Tables and one based on Merkle Hash Trees.
We show that the two schemes are secure and privacy-preserving in the random oracle model.

\end{abstract}


\section{Introduction}

\noindent Storage systems allow everyone to upload his/her data on cloud servers, 
and thus avoid keeping them on his/her own devices that have often limited storage capacity and power.

Nevertheless, storage services are susceptible to attacks or failures, and lead to possible non-retrievable losses of the stored data.
Indeed, storage systems are vulnerable to internal and external attacks that harm the data integrity
even being more powerful and reliable than the data owner's personal computing devices. 
A solution is to construct a system that offers an efficient, frequent and secure data integrity check process to the data owner
such that the frequency of data integrity verification and the percentage of audited data should not be limited 
by computational and communication costs on both cloud server's and data owner's sides.
\\

A Provable Data Possession (PDP) enables a data owner, called the \textit{client}, to verify the integrity of his/her data stored on an untrusted cloud server, 
without having to retrieve them.
Informally, the client first divides his/her data into blocks, generates tags on each block, and then forwards all these elements to the server.
In order to check whether the data are correctly stored by the server, 
the client sends a challenge such that the server replies back by creating a proof of data possession.
If the proof is correct, then this means that the storage of the data is correctly done by the server; otherwise, this means that the server is actually cheating somehow.

Natural extension features of PDP include: 
\begin{enumerate}
\item Dynamicity (D) that enables the client to update his/her data stored on the server via three operations (insertion, deletion and modification);
\item Public verifiability (PV) that allows a client to indirectly check that the server correctly stores his/her data by enabling a Third Party Auditor (TPA) or everyone else to do the audit;
\item Data privacy (DP) preservation that ensures that the contents of the stored data are not leaked to neither the TPA nor anyone else.
\end{enumerate}

We require that a Dynamic PDP (DPDP) with PV and DP system is secure 
at untrusted server, which means that the server cannot successfully generate a proof of data possession that is correct without actually storing all the data.
In addition, 
a DPDP with PV and DP system should be data privacy-preserving, 
which means that the TPA should not learn anything about the client's data even by having access to the public information.
\\

Gritti et al. \cite{GSP15} recently constructed an efficient and practical DPDP system with PV and DP.
However, we have found three attacks threatening this construction:
\begin{enumerate}
\item The \textit{replace attack} enables the server to store only one block of a file $m$ and still pass the data integrity verification on any number of blocks;
\item The \textit{replay attack} permits the server to keep the old version of a block $m_{i}$ 
and the corresponding tag $T_{m_{i}}$, after the client asked to modify them by sending the new version of these elements, 
and still pass the data integrity verification;
\item The \textit{attack against data privacy} allows the TPA to distinguish files when proceeding the data integrity check without accessing their contents. 
\end{enumerate}



We then propose two solutions to overcome the adversarial issues threatening the DPDP scheme with PV and DP in \cite{GSP15}.
We give a first new publicly verifiable DPDP construction based on Index Hash Tables (IHT) in the random oracle model. 
 We prove that such scheme is secure against replace and replay attacks
 as well as is data privacy-preserving according to a model differing from the one proposed in \cite{GSP15}.
 We present a second new publicly verifiable DPDP construction based on Merkle Hash Trees (MHT) in the random oracle model. 
 We demonstrate that such scheme is not vulnerable against the three attacks mentioned above. 
 In particular, we use the existing model given in \cite{GSP15} to prove that the MHT-based scheme is data privacy-preserving. 


\subsection{Related Work}

\noindent Ateniese et al. \cite{ABCHKPS07} introduced the notion of Provable Data Possession (PDP)
which allows a client to verify the integrity of his/her data stored at an untrusted server without retrieving the entire file. 
Their scheme is designed for static data and used homomorphic authenticators as tags based on public key encryption for auditing the data file. 
Subsequently, Ateniese et al. \cite{APMT08} improved the efficiency of the aforementioned PDP scheme by using symmetric keys. 
The resulting scheme gets lower overhead and partially supports partial dynamic data operations. 
Thereafter, various PDP constructions were proposed in the literature \cite{WWRL10,YWRL10,HZY11,YAMTRSD14}. 
Moreover, PDP schemes with the property of full dynamicity were suggested in \cite{EKPT09,ZWHAHY11,ZAHYAH13,WWRL09,LM12,WWRCL12}.
An extension of DPDP includes version control \cite{EK13,CC14} where all data changes are recorded into a repository and
any version of the data can be retrieved at any time.
DPDP protocols with multi-update capability were suggested in \cite{EKBKO13,EKO14}.
More recently, data privacy-preserving and publicly verifiable PDP schemes were presented in \cite{WLL12,WLL12-2,WCWRL13,FYMY15,WLL15,GSP15}.

A similar concept to PDP, called Proof of Retrievability (POR), was first defined by Juels and Kaliski \cite{JK07}.
It allows a client to verify the integrity of his/her data stored at an untrusted server, to correct the possible errors and to retrieve the entire file. 
Challenge requests are limited and require randomly-valued check blocks that are called sentinels and added to the file.
Thereafter, Shacham and Waters \cite{SW08} proposed two POR schemes built on BLS signatures and pseudo-random functions.
They achieved unlimited number of challenge requests and public verifiability, and reduced communication overhead.
Subsequent works on distributed systems followed in \cite{BJO09,BJO09-2,DVW09}, as well as POR protocols with data dynamicity in \cite{WWLRL09,SSP13}.



\section{Preliminaries}
\label{prelim}

\subsection{Bilinear Maps}

Let $\mathbb{G}_{1}$, $\mathbb{G}_{2}$ and $\mathbb{G}_{T}$ be three multiplicative cyclic groups of prime order
$p \in \Theta(2^{\lambda})$ (where $\lambda$ is the security parameter).
Let $g_{1}$ be a generator of $\mathbb{G}_{1}$, $g_{2}$ be a generator of $\mathbb{G}_{2}$ that
we denote $<g_{1}>=\mathbb{G}_{1}$ and $<g_{2}>=\mathbb{G}_{2}$.
Let $e : \mathbb{G}_{1} \times \mathbb{G}_{2} \to \mathbb{G}_{T}$ be a bilinear map with the following properties:
\begin{enumerate}
 \item \textit{Bilinearity:} $\forall u \in \mathbb{G}_{1}, \forall v \in \mathbb{G}_{2}, \forall a,b \in \mathbb{Z}_{p},$

$e(u^{a},v^{b})=e(u,v)^{ab}$,
 \item \textit{Non-degeneracy:} $e(g_{1},g_{2}) \neq 1_{\mathbb{G}_{T}}$.
\end{enumerate}
$\mathbb{G}_{1}$ and $\mathbb{G}_{2}$ are said to be bilinear groups if the group operation in $\mathbb{G}_{1} \times \mathbb{G}_{2}$ and the bilinear map $e$ are both efficiently computable.
We can easily see that $e$ is symmetric since $e(g_{1}^{a},g_{2}^{b})=e(g_{1},g_{2})^{ab}=e(g_{1}^{b},g_{2}^{a})$.
Let \textsf{GroupGen} denote an algorithm that on input the security parameter $\lambda$,
outputs the parameters $(p,\mathbb{G}_{1},\mathbb{G}_{2},\mathbb{G}_{T},e,g_{1},g_{2})$ as defined above.

\subsection{Assumptions}

\subsubsection{Discrete Logarithm}

The Discrete Logarithm (DL) problem is as follows.
Let $\mathbb{G}_{1}$ be a group of prime order $p$ according to the security parameter $\lambda$.
Let $a \in_{R} \mathbb{Z}_{p}$ and $<g_{1}>=\mathbb{G}_{1}$.
If $\mathcal{A}$ is given an instance $(  g_{1},g_{1}^{a})$,
it remains hard to extract $a \in \mathbb{Z}_{p}$.

The DL assumption holds if no polynomial-time adversary $\mathcal{A}$ has non-negligible advantage in solving the DL problem.

\subsubsection{Computational Diffie-Hellman}

The Computational Diffie-Hellman (CDH) problem is as follows.
Let $\mathbb{G}$ be a group of prime order $p$ according to the security parameter $\lambda$.
Let $a,b \in_{R} \mathbb{Z}_{p}$ and $<g>=\mathbb{G}$.
If $\mathcal{A}$ is given an instance $(  g,g^{a},g^{b})$,
it remains hard to compute $g^{ab} \in \mathbb{G}$.

The CDH assumption holds if no polynomial-time adversary $\mathcal{A}$ has non-negligible advantage in solving the CDH problem.

\subsubsection{Decisional Diffie-Hellman Exponent}

The Decisional Diffie-Hellman Exponent ($(s+1)$-DDHE) problem is as follows.
Let $\mathbb{G}_{1}$ and $\mathbb{G}_{2}$ be two groups of prime order $p$ according to the security parameter $\lambda$.
Let $\beta \in_{R} \mathbb{Z}_{p}$, $<g_{1}>=\mathbb{G}_{1}$ and $<g_{2}>=\mathbb{G}_{2}$.
If $\mathcal{A}$ is given an instance $(g_{1},g_{1}^{\beta},\cdots,g_{1}^{\beta^{s+1}},g_{2},g_{2}^{\beta},Z)$,
it remains hard to decide if either $Z= g_{1}^{\beta^{s+2}}$ or $Z$ is a random element in $\mathbb{G}_{1}$.

The $(s+1)$-DDHE assumption holds if no polynomial-time adversary $\mathcal{A}$ has non-negligible advantage in solving the $(s+1)$-DDHE problem.

\subsection{Definition of DPDP Scheme with PV and DP}

\noindent Let $m$ be a data file to be stored that is divided into $n$ \textit{blocks} $m_{i}$,
and then each block $m_{i}$ is divided into $s$ \textit{sectors} $m_{i,j} \in \mathbb{Z}_{p}$, where $p$ is a large prime.


A DPDP scheme with PV and DP is made of the following algorithms:

 \noindent $\bullet$ $\textsf{KeyGen}(\lambda) \to (pk,sk)$.
 On input the security parameter $\lambda$, output a pair of public and secret keys $(pk,sk)$.
 
 \noindent $\bullet$ $\textsf{TagGen}(pk,sk,m_{i}) \to T_{m_{i}}$. 
\textsf{TagGen} is independently run for each block. Therefore, to generate the tag $T_{m}$ for a file $m$,
\textsf{TagGen} is run $n$ times.
 On inputs the public key $pk$, the secret key $sk$ and a file $m=(m_{1},\cdots,m_{n})$, output a tag $T_{m}=(T_{m_{1}},\cdots,T_{m_{n}})$ 
 where each block $m_{i}$ has its own tag $T_{m_{i}}$.
 The client sets all the blocks $m_{i}$ in an ordered collection $\mathbb{F}$ 
 and all the corresponding tags $T_{m_{i}}$ in an ordered collection $\mathbb{E}$.
 He/she sends $\mathbb{F}$ and $\mathbb{E}$ to the server and removes them from his/her local storage.
 
\noindent $\bullet$ $\textsf{PerfOp}(pk,\mathbb{F},\mathbb{E},info=(\mbox{operation},l,m_{l},T_{m_{l}}))$ 

\noindent $ \to (\mathbb{F'},\mathbb{E}',\nu')$.
 On inputs the public key $pk$, the previous collection $\mathbb{F}$ of all the blocks, 
 the previous collection $\mathbb{E}$ of all the corresponding tags,
 the type of the data operation to be performed, 
the rank $l$ where the data operation is performed in $\mathbb{F}$,
the block $m_{l}$ to be updated 
and the corresponding tag $T_{m_{l}}$ to be updated, 
  output the updated block collection $\mathbb{F}'$, 
 the updated tag collection $\mathbb{E}'$ 
 and an updating proof $\nu'$. 
 
 For the operation:
\begin{enumerate}
\item \textit{Insertion:} $m_{l} =m_{\frac{i_{1} + i_{2}}{2}}$ is inserted between the consecutive blocks $m_{i_{1}}$ and $m_{i_{2}}$ and 
 $T_{m_{l}}=T_{m_{\frac{i_{1} + i_{2}}{2}}}$ is inserted between the consecutive tags $T_{m_{i_{1}}}$ and $T_{m_{i_{2}}}$.
  We assume that $m_{\frac{i_{1} + i_{2}}{2}}$ and $T_{m_{\frac{i_{1} + i_{2}}{2}}}$ were provided by the client to the server, 
 such that $T_{m_{\frac{i_{1} + i_{2}}{2}}}$ was correctly computed by running \textsf{TagGen}.
%
\item \textit{Deletion:} $m_{l} =m_{i}$ is deleted, meaning that $m_{i_{1}}$ is followed by $m_{i_{2}}$ and $T_{m_{l}}=T_{m_{i}}$ is deleted, 
 meaning that $T_{m_{i_{1}}}$ is followed by $T_{m_{i_{2}}}$, such that $i_{1},i,i_{2}$ were three consecutive ranks.
\item \textit{Modification:} $m_{l} =m_{i}'$ replaces $m_{i}$ and $T_{m_{l}}=T_{m_{i}'}$ replaces $T_{m_{i}}$.
 We assume that $m_{i}'$ and $T_{m_{i}'}$ were provided by the client to the server, 
 such that $T_{m_{i}'}$ was correctly computed by running \textsf{TagGen}.
\end{enumerate}

  \noindent $\bullet$ $\textsf{CheckOp}(pk,\nu') \to 0/1$. 
 On inputs the public key $pk$ and the updating proof $\nu'$ sent by the server, output $1$ if $\nu'$ is a correct updating proof; output $0$ otherwise.
  
  \noindent $\bullet$ $\textsf{GenProof}(pk,F,chal,\Sigma) \to \nu$.  
 On inputs the public key $pk$, an ordered collection $F \subset \mathbb{F}$ of blocks, a challenge $chal$ and an ordered collection $\Sigma \subset \mathbb{E}$ 
 which are the tags corresponding to the blocks in $F$, 
 output a proof of data possession $\nu$ for the blocks in $F$ that are determined by $chal$. 
 
 \noindent $\bullet$ $\textsf{CheckProof}(pk,chal,\nu) \to 0/1$. 
 On inputs the public key $pk$, the challenge $chal$ and the proof of data possession $\nu$, output $1$ 
 if $\nu$ is a correct proof of data possession for the blocks determined by $chal$; output $0$ otherwise.

 \paragraph{Correctness.}
We require that a DPDP with PV and DP is \textit{correct} if
for $(pk,sk) \gets \textsf{KeyGen}(\lambda)$, $T_{m} \gets \textsf{TagGen}(pk,sk,$ $m)$, 
$(\mathbb{F'},\mathbb{E}',\nu') \gets \textsf{PerfOp}$ $(pk,\mathbb{F},\mathbb{E},info)$,
$\nu \gets \textsf{GenProof}(pk,$ $F,chal,\Sigma)$,
then $1 \gets \textsf{CheckOp}(pk,$ $\nu')$ and $1 \gets \textsf{CheckProof}(pk,$ $chal,\nu)$.

\paragraph{N.B.}
 The set of ranks is $[1,n]$ at the first upload; it then becomes $(0,n+1) \cap \mathbb{Q}$ after operations as in the construction in \cite{GSP15}.

\subsection{Security and Privacy Models}

\subsubsection{Security Model against the Server}
\label{secagservoriginal}

\noindent This model against the server is given in \cite{GSP15}, and follows the one proposed in \cite{ABCHKPS07,EKPT09}. 

 We consider a DPDP with PV and DP as defined above.
 Let a data possession game between a challenger $\mathcal{B}$ and an adversary $\mathcal{A}$ (acting as the server) be as follows:
 
  \noindent $\diamond$
  \textit{Setup.}
  $\mathcal{B}$ runs $(pk,sk) \gets \textsf{KeyGen}(\lambda)$ such that $pk$ is given to $\mathcal{A}$ while $sk$ is kept secret.
  
   \noindent $\diamond$
  \textit{Adaptive Queries.} 
  First, $\mathcal{A}$ is given access to a tag generation oracle $\mathcal{O}_{TG}$.
  $\mathcal{A}$ chooses blocks $m_{i}$ and gives them to $\mathcal{B}$, for $i \in [1,n]$.
  $\mathcal{B}$ runs $\textsf{TagGen}(pk,sk,m_{i}) \to T_{m_{i}}$ and gives them to $\mathcal{A}$.
  Then, $\mathcal{A}$ creates two ordered collections $\mathbb{F}=\{m_{i}\}_{i \in [1,n]}$ of blocks and
  $\mathbb{E} =\{T_{m_{i}}\}_{i \in [1,n]}$ of the corresponding tags.
  
  Then, $\mathcal{A}$ is given access to a data operation performance oracle $\mathcal{O}_{DOP}$.
  For $i \in [1,n]$, $\mathcal{A}$ gives to $\mathcal{B}$ a block $m_{i}$ and $info_{i}$ about the operation that $\mathcal{A}$ wants to perform.
  $\mathcal{A}$ also submits two new ordered collections $\mathbb{F}'$ of blocks and $\mathbb{E}'$ of tags, and the updating proof $\nu'$. 
  $\mathcal{B}$ runs $\textsf{CheckOp}(pk,\nu')$ and replies the answer to $\mathcal{A}$.
  If the answer is $0$, then $\mathcal{B}$ aborts; otherwise, it proceeds.
  The above interaction between $\mathcal{A}$ and $\mathcal{B}$ can be repeated.
Note that the set of ranks has changed after calls to the oracle $\mathcal{O}_{DOP}$.
  
 \noindent $\diamond$
  \textit{Challenge.} 
  $\mathcal{A}$ chooses blocks $m_{i}^{*}$ and $info_{i}^{*}$, 
  for $i \in \mathcal{I} \subseteq (0,n+1) \cap \mathbb{Q}$.
  Adaptive queries can be again made by $\mathcal{A}$, such that the first $info_{i}^{*}$ specifies a full re-write update 
  (this corresponds to the first time that the client sends a file to the server). $\mathcal{B}$ still checks the data operations.
  
  For $i \in \mathcal{I}$, the final version of 
  $m_{i}$ is considered such that these blocks were created regarding the operations requested by $\mathcal{A}$, and verified and accepted by $\mathcal{B}$ beforehand.
  $\mathcal{B}$ sets $\mathbb{F} = \{m_{i}\}_{i \in \mathcal{I}}$ of these blocks 
  and $\mathbb{E}= \{T_{m_{i}}\}_{i \in \mathcal{I}}$ of the corresponding tags.
 It then sets two ordered collections $F=\{m_{i_{j}}\}_{i_{j} \in \mathcal{I},j \in [1,k]} \subset \mathbb{F}$ 
 and $\Sigma=\{T_{m_{i_{j}}}\}_{i_{j} \in \mathcal{I},j \in [1,k]} \subset \mathbb{E}$.
  It computes a resulting challenge $chal$ for $F$ and $\Sigma$ and sends it to $\mathcal{A}$.
  
  \noindent $\diamond$
 \textit{Forgery.}
 $\mathcal{A}$ computes a proof of data possession $\nu^{*}$ on $chal$.
  Then, $\mathcal{B}$ runs $\textsf{CheckProof}(pk,$ $chal,\nu^{*})$ and replies the answer to $\mathcal{A}$.
  If the answer is $1$ then $\mathcal{A}$ wins.

  The advantage of $\mathcal{A}$ in winning the data possession game is defined as $Adv_{\mathcal{A}}(\lambda) = Pr [ \mathcal{A} $ $ \mbox{ wins}]$.
  The DPDP with PV and DP is \textit{secure against the server} if there is no PPT (probabilistic polynomial-time) adversary $\mathcal{A}$ who can win the above game 
  with non-negligible advantage $Adv_{\mathcal{A}}(\lambda)$.
 

\subsubsection{Data Privacy Model against the TPA}
\label{dpsm}

In a DPDP protocol, we aim to ensure that data privacy is preserved at the verification step, meaning that data are accessible to all but protected only via a non-cryptographic access control,
and the verification process does not leak any information on the data blocks.

\paragraph{First Data Privacy Model.}
This model is found in the literature \cite{WWRL10,WCWRL13} to show that public auditing systems preserve data privacy.

We consider a DPDP with PV and DP as defined above.
Let the first data privacy game between a challenger $\mathcal{B}$ and an adversary $\mathcal{A}$ (acting as the TPA) be as follows:

  \noindent $\diamond$
\textit{Setup.} $\mathcal{B}$ runs ${\sf KeyGen}(\lambda)$ to generate $(pk,sk)$ and gives $pk$ to $\mathcal{A}$,
while $sk$ is kept secret.

  \noindent $\diamond$
\textit{Queries.} $\mathcal{A}$ is allowed to make queries as follows.
$\mathcal{A}$ sends a file $m = (m_{1},\cdots,m_{n})$ to $\mathcal{B}$.
$\mathcal{B}$ computes $T_{m} = (T_{m_{1}},\cdots,T_{m_{n}})$ and gives it back to $\mathcal{A}$.
Then, two ordered collections $\mathbb{F}= \{m_{i}\}_{i \in [1,n]}$ of blocks and $\mathbb{E} = \{T_{m_{i}}\}_{i \in [1,n]}$ of tags are created.

  \noindent $\diamond$
\textit{Challenge.} $\mathcal{A}$ submits a challenge $chal$ containing $k \le n$ ranks, the $k$ corresponding blocks in $F$
and their $k$ tags in $\Sigma$.
  
 \noindent $\diamond$
\textit{Generation of the Proof.} $\mathcal{B}$ computes a proof of data possession $\nu^{*} \gets \textsf{GenProof}(pk,F,chal,\Sigma)$
such that the blocks in $F$ are determined by $chal$ and $\Sigma$ contains the corresponding tags.

$\mathcal{A}$ succeeds in the first data privacy game if $F \nsubseteq \mathbb{F}$ and $\Sigma \nsubseteq \mathbb{E}$, and
$\textsf{CheckProof} (pk,chal,\nu^{*})$ $ \to 1$.
The advantage of $\mathcal{A}$ in winning the first data privacy game is defined as $Adv_{\mathcal{A}}(\lambda) = Pr[ \mathcal{A} \mbox{ succeeds}]$.

The DPDP with PV and DP is \textit{data privacy-preserving} 
if there is no PPT adversary $\mathcal{A}$ who can win the above game with non-negligible advantage $Adv_{\mathcal{A}}(\lambda)$. 
This implies that there is no $\mathcal{A}$ who can recover the file from a given tag tuple with non-negligible probability.

 \paragraph{Second Data Privacy Model.}
This model is given in \cite{GSP15}, and follows the one proposed in \cite{FYMY15,YAMTRSD14}. 
Observe that such model is based on indistinguishability.
 
 We consider a DPDP with PV and DP as defined above.
 Let a second data privacy game between a challenger $\mathcal{B}$ and an adversary $\mathcal{A}$ (acting as the TPA) be as follows:
 
\noindent $\diamond$
  \textit{Setup.} $\mathcal{B}$ runs $\textsf{KeyGen}(\lambda)$ to generate $(pk,sk)$ and gives $pk$ to $\mathcal{A}$,
 while $sk$ is kept secret.
  
  \noindent $\diamond$
  \textit{Queries.} $\mathcal{A}$ is allowed to make queries as follows.
  $\mathcal{A}$ sends a file $m$ to $\mathcal{B}$.
  $\mathcal{B}$ computes the corresponding $T_{m}$ and gives it to $\mathcal{A}$.
  
\noindent $\diamond$
  \textit{Challenge.} $\mathcal{A}$ submits two different files $m_{0}$ and $m_{1}$ of equal length, such that they have not be chosen in the phase Queries, and sends
  them to $\mathcal{B}$. 
  $\mathcal{B}$ generates $T_{m_{0}}$ and $T_{m_{1}}$ by running \textsf{TagGen}, randomly chooses a bit $b \in_{R} \{0,1\}$ and forwards $T_{m_{b}}$ to $\mathcal{A}$.
  Then, $\mathcal{A}$ sets a challenge $chal$ and sends it to $\mathcal{B}$.
  $\mathcal{B}$ generates a proof of data possession $\nu^{*}$ based on $m_{b}$, $T_{m_{b}}$ and $chal$, and replies to $\mathcal{A}$ by giving $\nu^{*}$.
  
\noindent $\diamond$
  \textit{Guess.} Finally, $\mathcal{A}$ chooses a bit $b' \in \{0,1\}$ and wins the game if $b'=b$.
 
 The advantage of $\mathcal{A}$ in winning the second data privacy game 
 is defined as $Adv_{\mathcal{A}}(\lambda)= |Pr[b'=b] - \frac{1}{2} |$.
 
The DPDP with PV and DP is \textit{data privacy-preserving} 
if there is no PPT adversary $\mathcal{A}$ who can win the above game with non-negligible advantage $Adv_{\mathcal{A}}(\lambda)$.

%
%
%
%

\section{The Three Attacks}
\label{attacks}


\subsection{DPDP construction with PV and DP in \cite{GSP15}}
\label{schemedef}

\noindent 
The DPDP scheme with PV and DP construction presented in \cite{GSP15} is as follows:

\noindent $\bullet$ $\textsf{KeyGen}(\lambda) \to (pk,sk)$.
  The client runs $\textsf{Group-}$ $\textsf{Gen}(\lambda) \to (p,\mathbb{G}_{1},\mathbb{G}_{2},\mathbb{G}_{T},e,g_{1},g_{2})$
  such that on input the security parameter $\lambda$, \textsf{GroupGen}
generates the cyclic groups $\mathbb{G}_{1}$, $\mathbb{G}_{2}$ and $\mathbb{G}_{T}$ of prime order $p = p(\lambda)$ with the
bilinear map $e : \mathbb{G}_{1} \times \mathbb{G}_{2} \to \mathbb{G}_{T}$. 
Let $<g_{1}>=\mathbb{G}_{1}$ and $<g_{2}>=\mathbb{G}_{2}$.
 Then, $h_{1},\cdots,h_{s} \in_{R} \mathbb{G}_{1}$ and $a \in_{R} \mathbb{Z}_{p}$ are randomly chosen.
 Finally, he/she sets the public key $pk= (p,\mathbb{G}_{1},\mathbb{G}_{2},\mathbb{G}_{T},e,g_{1},g_{2},h_{1},\cdots,h_{s},$ $g_{2}^{a})$ 
 and the secret key $sk=a$.
 
 \noindent $\bullet$ $\textsf{TagGen}(pk,sk,m_{i}) \to T_{m_{i}}$. 
 A file $m$ is split into $n$ blocks $m_{i}$, for $i \in [1,n]$.
Each block $m_{i}$ is then split into $s$ sectors $m_{i,j} \in \mathbb{Z}_{p}$, for $j \in [1,s]$.
Therefore, the file $m$ can be seen as a $n \times s$ matrix with elements denoted as $m_{i,j}$.
The client computes
$ T_{m_{i}} =  (\prod_{j=1}^{s} h_{j}^{m_{i,j}})^{-sk} =\prod_{j=1}^{s} h_{j}^{- a \cdot m_{i,j}}$.
Yet, he/she sets $T_{m}= (T_{m_{1}},\cdots,$ $T_{m_{n}}) \in \mathbb{G}_{1}^{n}$.
%

\noindent $\bullet$ $\textsf{PerfOp}(pk,\mathbb{F},\mathbb{E},info=(\mbox{operation},l,m_{l},T_{m_{l}}))$ $ \to (\mathbb{F'},\mathbb{E}',\nu')$.
The server first selects at random $u_{j} \in_{R} \mathbb{Z}_{p}$, for $j \in [1,s]$, and computes $U_{j}=h_{j}^{u_{j}}$.
It also chooses at random $w_{l} \in_{R} \mathbb{Z}_{p}$ and sets $c_{j}= m_{l,j} \cdot w_{l}  + u_{j} $, 
$C_{j}=h_{j}^{c_{j}}$, and $d= T_{m_{l}}^{w_{l}}$.
 Finally, it returns $\nu' = (U_{1},\cdots,U_{s},C_{1},\cdots,C_{s},d,$ $w_{l}) \in \mathbb{G}_{1}^{2s+1}$ to the TPA.
  
  For the operation:
\begin{enumerate}
\item \textit{Insertion:} $(l,m_{l},T_{m_{l}}) = (\frac{i_{1}+i_{2}}{2},m_{\frac{i_{1}+i_{2}}{2}},$ 

$T_{m_{\frac{i_{1}+i_{2}}{2}}})$;
\item \textit{Deletion:} $(l,m_{l},T_{m_{l}}) = (i,\_,\_)$, meaning that $m_{l}$ and $T_{m_{l}}$ are not required
  (the server uses $m_{i}$ and $T_{m_{i}}$ that are kept on its storage to generate $\nu'$);
\item \textit{Modification:} $(l,m_{l},T_{m_{l}}) =(i,m_{i}',T_{m_{i}'})$.
\end{enumerate}

\noindent $\bullet$ $\textsf{CheckOp}(pk,\nu') \to 0/1$.
 The TPA has to check whether the following equation holds:
    \begin{eqnarray}
 e(d,g_{2}^{a}) \cdot e(\prod_{j=1}^{s} U_{j} ,g_{2}) &\stackrel{?}{=}&  e(\prod_{j=1}^{s} C_{j},g_{2})
  \label{eqt15-basic}
  \end{eqnarray}
    If Eq. \ref{eqt15-basic} holds, then the TPA returns $1$ to the client; otherwise, it returns $0$ to the client.
   
 \noindent  $\bullet$ $\textsf{GenProof}(pk,F,chal,\Sigma) \to \nu$.
 The TPA first chooses $I \subseteq (0,n+1) \cap \mathbb{Q}$, randomly chooses $|I|$ elements $v_{i} \in_{R} \mathbb{Z}_{p}$ and sets $chal=\{(i,v_{i})\}_{i \in I}$.
 
 After receiving $chal$, the server sets $F = \{m_{i}\}_{i \in I} \subset \mathbb{F}$ of blocks
 and $\Sigma = \{T_{m_{i}}\}_{i \in I} \subset \mathbb{E}$ which are the tags corresponding to the blocks in $F$.
 It then selects at random $r_{j} \in_{R} \mathbb{Z}_{p}$, for $j \in [1,j]$, and computes $R_{j}=h_{j}^{r_{j}}$.
 It also sets $b_{j}=\sum_{(i,v_{i}) \in chal} m_{i,j} \cdot v_{i}  + r_{j}$, $B_{j}=h_{j}^{b_{j}}$ for $j \in [1,s]$, 
 and $c =\prod_{(i,v_{i}) \in chal} T_{m_{i}}^{v_{i}}$.
 Finally, it returns $\nu = (R_{1},\cdots,R_{s},B_{1},\cdots,B_{s},c) \in \mathbb{G}_{1}^{2s+1}$ to the TPA.
  
 \noindent $\bullet$ $\textsf{CheckProof}(pk,chal,\nu) \to 0/1$.
  The TPA has to check whether the following equation holds:
  \begin{eqnarray}
    e(c,g_{2}^{a}) \cdot e(\prod_{j=1}^{s} R_{j} ,g_{2})  &\stackrel{?}{=}&   e(\prod_{j=1}^{s} B_{j},g_{2})
  \label{eqt25-basic}
  \end{eqnarray}
   If Eq. \ref{eqt25-basic} holds, then the TPA returns $1$ to the client; otherwise, it returns $0$ to the client.

 \paragraph{Correctness.}
Given the proof of data possession $\nu$ and the updating proof $\nu'$, we have:
\begin{eqnarray*}
 e(c,g_{2}^{a}) \cdot e(\prod_{j=1}^{s} R_{j} ,g_{2}) 
  &=& e ( \prod_{\substack{(i,v_{i}) \\ \in chal}} T_{m_{i}}^{v_{i}},g_{2}^{a} ) \cdot e(\prod_{j=1}^{s} h_{j}^{r_{j}} ,g_{2}) \\
 &=&  e(\prod_{j=1}^{s} h_{j}^{b_{j}},g_{2}) =e(\prod_{j=1}^{s} B_{j},g_{2}) 
\end{eqnarray*}

\begin{eqnarray*}
  e(d,g_{2}^{a})  \cdot e(\prod_{j=1}^{s} U_{j} ,g_{2}) 
  &=&  e( T_{m_{i}}^{w_{i}},g_{2}^{a}) \cdot e(\prod_{j=1}^{s} h_{j}^{u_{j}} ,g_{2}) \\
 &=& e(\prod_{j=1}^{s} h_{j}^{c_{j}},g_{2}) = e(\prod_{j=1}^{s} C_{j},g_{2})
\end{eqnarray*}

\paragraph{N.B.}
In the construction in \cite{GSP15}, 
the definition of the tag $T_{m_{i}}$ corresponding to the block $m_{i}$ and enabling to remotely verify the data integrity
is independent of the rank $i$; thus, this begs for being used for an attack.
Note that if $m_{i}=0$, then $T_{m_{i}}=1$ and thus, one can trivially cheat since the tag is independent of the file.

\subsection{Replace Attack}
\label{replaceattack}

\noindent 
Let the server store only one block (e.g. $m_{1}$) instead of $n$ blocks as the client believes.
The TPA audits the server by sending it a challenge $chal$ for blocks with ranks in $I \subseteq [1,n]$ such that $|I| \le n$.
The server generates a proof of data possession on the $|I| $ blocks $m_{1}$ (instead of the blocks defined by $chal$) by using 
$|I| $ times the block $m_{1}$ to obtain the proof of data possession. 
The attack is successful if the server manages to pass the verification process and has its proof of data possession being accepted by the TPA.
\\

The client computes $T_{m} = (T_{m_{1}},\cdots,T_{m_{n}}) \in \mathbb{G}_{1}^{n}$ for a file $m = (m_{1},\cdots,m_{n})$
where $T_{m_{i}} = (\prod_{j=1}^{s} h_{j}^{m_{i,j}})^{-sk} =  (\prod_{j=1}^{s} h_{j}^{m_{i,j}})^{-a} $
for $s$ public elements $h_{j} \in \mathbb{G}_{1}$ and the secret key $sk=a \in \mathbb{Z}_{p}$.
Then, the client stores all the blocks $m_{i}$ in $\mathbb{F}$ 
and the tags $T_{m_{i}}$ in $\mathbb{E}$, forwards these collections to the server and deletes them from his/her local storage.
 
Yet, the server is asked to generate a proof of data possession $\nu$. 
We assume that it only stores $m_{1}$ while it has deleted $m_{2},\cdots,m_{n}$ and we show that it can still pass the verification process.
The TPA prepares a challenge $chal$ by choosing a set $I \subseteq [1,n]$ (without loss of generality, we assume that the client has not requested the server for data operations yet).
The TPA then randomly chooses $|I|$ elements $v_{i} \in_{R} \mathbb{Z}_{p}$ and sets $chal=\{(i,v_{i})\}_{i \in I}$.
Second, after receiving $chal$, the server sets $F = \{m_{1}\}_{i \in I}$ $ \subset \mathbb{F}$ of blocks (instead of $F = \{m_{i}\}_{i \in I}$) 
and $\Sigma = \{T_{m_{1}}\}_{i \in I} \subset \mathbb{E}$ 
(instead of $\Sigma = \{T_{m_{i}}\}_{i \in I}$).
The server finally forwards $\nu = (R_{1},\cdots,R_{s},B_{1},$ $\cdots,B_{s},c) \in \mathbb{G}_{1}^{2s+1}$ to the TPA,
where $R_{j}=h_{1}^{r_{j}}$ for $r_{j} \in_{R} \mathbb{Z}_{p}$ and $B_{j} = h_{j}^{\sum_{(i,v_{i}) \in chal} m_{1,j} \cdot v_{i}  + r_{j}}$
(instead of $B_{j}=h_{j}^{\sum_{(i,v_{i}) \in chal} m_{i,j} \cdot v_{i}  + r_{j}}$) for $j \in [1,s]$, and
$c =\prod_{(i,v_{i}) \in chal} T_{m_{1}}^{v_{i}}$ (instead of $c =\prod_{(i,v_{i}) \in chal} T_{m_{i}}^{v_{i}}$).

The TPA has to check whether the following equation holds:
 \begin{eqnarray}
 e(c,g_{2}^{a}) \cdot e(\prod_{j=1}^{s} R_{j} ,g_{2}) &\stackrel{?}{=}& e(\prod_{j=1}^{s} B_{j},g_{2})
 \label{eqt3-replace}
 \end{eqnarray}
  If Eq. \ref{eqt3-replace} holds, then the TPA returns $1$ to the client; otherwise, it returns $0$ to the client.

  \paragraph{Correctness.}
  Given the proof of data possession $\nu$, we have:
\begin{eqnarray*}
   e(c,g_{2}^{a}) \cdot e(\prod_{j=1}^{s} R_{j} ,g_{2}) &=&  e ( \prod_{(i,v_{i})  \in chal} T_{m_{1}}^{v_{i}},g_{2}^{a} ) \cdot e(\prod_{j=1}^{s} h_{j}^{r_{j}} ,g_{2}) \\
 &=& e ( \prod_{(i,v_{i})  \in chal} \prod_{j=1}^{s} h_{j}^{m_{1,j} \cdot (-a) \cdot v_{i}},g_{2}^{a} ) \cdot e(\prod_{j=1}^{s} h_{j}^{r_{j}} ,g_{2}) \\
 &=&  e(\prod_{j=1}^{s} h_{j}^{b_{j}},g_{2}) =e(\prod_{j=1}^{s} B_{j},g_{2}) 
\end{eqnarray*}
Therefore, Eq. \ref{eqt3-replace} holds, although the server is actually storing one block only.

 \paragraph{N.B.}
This attack is not due to the dynamicity property of the scheme in \cite{GSP15}. Such attack could happen even on static data.

\subsection{Replay Attack}

\noindent 
The client asks the server to replace $m_{i}$ with $m_{i}'$.
However, the server does not proceed and keeps $m_{i}$ on its storage.
Then, the TPA has to check that the operation has been correctly done and asks the server for an updating proof $\nu'$.
The server generates it, but using $m_{i}$ instead of $m_{i}'$.
The attack is successful if the server manages to pass the verification process and has $\nu'$ being accepted by the TPA.
\\

A client asks the server to modify the block $m_{i}$ by sending $m_{i}'$ and $T_{m_{i}'}$. 
However, the server does not follow the client's request and decides to keep $m_{i}$ and $T_{m_{i}}$, 
and deletes $m_{i}'$ and $T_{m_{i}'}$.

  The server receives $i$, $m_{i}'$ and $T_{m_{i}'}$ from the client but deletes them, 
  and generates the updating proof $\nu' = (U_{1},\cdots,U_{s},C_{1},\cdots,C_{s},d) \in \mathbb{G}_{1}^{2s+1}$
  by using $m_{i}$ and $T_{m_{i}}$ such that
  $U_{j}=h_{1}^{u_{j}}$ where $u_{j} \in_{R} \mathbb{Z}_{p}$ and
  $C_{j} = h_{j}^{m_{i,j} \cdot w_{i}  + u_{j}  }$ (instead of $C_{j}= h_{j}^{ m_{i,j}' \cdot w_{i}  + u_{j} }$) for $j \in [1,s]$,
  and $d= T_{m_{i}}^{w_{i}}$ (instead of $d= T_{m_{i}'}^{w_{i}}$).
  It gives $\nu'$ to the TPA.

  The TPA has to check whether the following equation holds:
   \begin{eqnarray}
 e(d,g_{2}^{a}) \cdot e(\prod_{j=1}^{s} U_{j} ,g_{2}) &\stackrel{?}{=}& e(\prod_{j=1}^{s} C_{j},g_{2})
 \label{eqt1-replay}
 \end{eqnarray}
   If Eq. \ref{eqt1-replay} holds, then the TPA returns $1$ to the client; otherwise, it returns $0$ to the client.
   
%

  \paragraph{Correctness.}
Given the updating proof $\nu'$, we have:
\begin{eqnarray*}
  e(d,g_{2}^{a})  \cdot e(\prod_{j=1}^{s} U_{j} ,g_{2}) &=&  e( T_{m_{i}}^{w_{i}},g_{2}^{a}) \cdot e(\prod_{j=1}^{s} h_{j}^{u_{j}} ,g_{2}) \\
 &=& e( \prod_{j=1}^{s} h_{j}^{m_{i,j} \cdot (-a) \cdot w_{i}},g_{2}^{a}) \cdot e(\prod_{j=1}^{s} h_{j}^{u_{j}} ,g_{2}) \\
 &=& e(\prod_{j=1}^{s} h_{j}^{c_{j}},g_{2}) = e(\prod_{j=1}^{s} C_{j},g_{2})
\end{eqnarray*}
Therefore, Eq. \ref{eqt1-replay} holds, although the server has not updated the block $m_{i}'$ and the corresponding tag $T_{m_{i}'}$.

 \paragraph{N.B.}
This attack is due to the dynamicity property of the scheme in \cite{GSP15}.

\subsection{Attack against Data Privacy }

\noindent 
The adversarial TPA and the server play the second data privacy game.
The TPA gives two equal-length blocks $m_{0}$ and $m_{1}$ to the server and the latter replies by sending $T_{m_{b}}$ of $m_{b}$ where $b \in_{R} \{0,1\}$ is a random bit.
Then, the TPA selects a bit $b' \in \{0,1\}$.
The attack is successful if using $m_{b'}$, the TPA can discover which block $m_{b} \in \{m_{0},m_{1}\}$ was chosen by the server.
\\

Let $m_{0} = (m_{0,1} , \cdots , m_{0,n})$ and $m_{1} =  (m_{1,1} ,$ $ \cdots , m_{1,n})$.
The server computes $T_{m_{b,i}} =(\prod_{j=1}^{s}$ $ h_{j}^{m_{b,i,j}})^{-sk} =(\prod_{j=1}^{s} h_{j}^{m_{b,i,j}})^{-a} $, for $b \in_{R} \{0,1\}$ and $i \in [1,n]$, 
and gives them to the TPA.
Note that 
$ e(T_{m_{b,i}},g_{2}) 
 = e( (\prod_{j=1}^{s} h_{j}^{m_{b,i,j}})^{-a},g_{2}) 
=e($ $ \prod_{j=1}^{s} h_{j}^{m_{b,i,j}},(g_{2}^{a})^{-1})$.
The computation of $e($ $\prod_{j=1}^{s} h_{j}^{m_{b,i,j}},(g_{2}^{a})^{-1})$ requires only public elements.
Therefore, for $b' \in \{0,1\}$, the TPA is able to generate the pairing $e(\prod_{j=1}^{s} h_{j}^{m_{b',i,j}},$ $(g_{2}^{a})^{-1})$
given $pk$ and the block that it gave to the server, and $e(T_{m_{b,i}},g_{2})$ given the tag sent by the server. Finally, the TPA compares them.
If these two pairings are equal, then $b'=b$; otherwise $b' \neq b$.

\paragraph{N.B.}
This attack is due to the public verifiability property of the scheme in \cite{GSP15} based on the definition of the second data privacy game. 

Moreover, in the proof for data privacy in \cite{GSP15}, the analysis is wrong:
the affirmation ``The probability $Pr[b'=b]$ must be equal to $\frac{1}{2}$ since the tags $T_{m_{b,i}}$, for $i \in [1,n]$,
and the proof $\nu^{*}$ are independent of the bit $b$.'' is incorrect since $T_{m_{b,i}}$ and $\nu^{*}$ actually depend on $b$.

\section{IHT-based DPDP scheme with PV and DP}
\label{ihtscheme}

\noindent A solution to avoid the replace attack is to embed the rank $i$ of $m_{i}$ into $T_{m_{i}}$. 
 When the TPA on behalf of the client checks $\nu$ generated by the server, it requires
to use all the ranks of the challenged blocks to process the verification. Such idea was proposed for the publicly verifiable scheme in \cite{SW08}.

A solution to avoid the replay attack is to embed the version number $vnb_{i}$ of $m_{i}$ into $T_{m_{i}}$.
The first time that the client sends $m_{i}$ to the server, $vnb_{i}=1$ (meaning that the first version of the block is uploaded) and is appended to $i$. 
When the client wants to modify $m_{i}$ with $m_{i}'$, he/she specifies $vnb_{i}=2$ (meaning that the second version of the block
is uploaded) and generates $T_{m_{i}'}$ accordingly. 
When the TPA on behalf of the client checks that the block was correctly updated by the server, it has to use both $i$ and $vnb_{i}$ of $m_{i}$.

Moreover, we stress that the rank $i$ of the block $m_{i}$ is unique.
More precisely, when a block is inserted, a new rank is created that has not been used and when a block is modified, the rank does not change. 
However, when a block is deleted, its rank does not disappear to ensure that it won't be used for another block and thus, to let the scheme remain secure.

\subsection{IHT-based Construction}

\noindent 
The IHT-based DPDP scheme with PV and DP construction is as follows:

\noindent $\bullet$ $\textsf{KeyGen}(\lambda) \to (pk,sk)$.
The client runs $\textsf{Group-}$  $\textsf{Gen}(\lambda) \to (p,\mathbb{G}_{1},\mathbb{G}_{2},\mathbb{G}_{T},e,g_{1},g_{2})$
  such that on input the security parameter $\lambda$, \textsf{GroupGen}
generates the cyclic groups $\mathbb{G}_{1}$, $\mathbb{G}_{2}$ and $\mathbb{G}_{T}$ of prime order $p = p(\lambda)$ with the
bilinear map $e : \mathbb{G}_{1} \times \mathbb{G}_{2} \to \mathbb{G}_{T}$. 
Let $<g_{1}>=\mathbb{G}_{1}$ and $<g_{2}>=\mathbb{G}_{2}$.
Let the hash function $H : \mathbb{Q} \times \mathbb{N} \to \mathbb{G}_{1}$ be a random oracle. 
 Then, $h_{1},\cdots,h_{s} \in_{R} \mathbb{G}_{1}$ and $a \in_{R} \mathbb{Z}_{p}$ are randomly chosen.
 Finally, he/she sets the public key $pk= (p,\mathbb{G}_{1},\mathbb{G}_{2},\mathbb{G}_{T},e,g_{1},g_{2},$ $h_{1},\cdots,h_{s},g_{2}^{a},H)$ 
 and the secret key $sk=a$.
 
 \noindent $\bullet$ $\textsf{TagGen}(pk,sk,m_{i}) \to T_{m_{i}}$. 
 A file $m$ is split into $n$ blocks $m_{i}$, for $i \in [1,n]$.
Each block $m_{i}$ is then split into $s$ sectors $m_{i,j} \in \mathbb{Z}_{p}$, for $j \in [1,s]$.
Therefore, the file $m$ can be seen as a $n \times s$ matrix with elements denoted as $m_{i,j}$.
The client computes
$ T_{m_{i}} = (H(i,vnb_{i}) \cdot \prod_{j=1}^{s} h_{j}^{m_{i,j}})^{-sk} 
=H(i,vnb_{i})^{-a} \cdot \prod_{j=1}^{s} h_{j}^{- a \cdot m_{i,j}}$.
Yet, he/she sets $T_{m}= (T_{m_{1}},\cdots,T_{m_{n}}) \in \mathbb{G}_{1}^{n}$.
%

\noindent $\bullet$ $\textsf{PerfOp}(pk,\mathbb{F},\mathbb{E},info=(\mbox{operation},l,m_{l},T_{m_{l}}))$

\noindent $\to (\mathbb{F'},\mathbb{E}',\nu')$.
The server first selects at random $u_{j} \in_{R} \mathbb{Z}_{p}$, for $j \in [1,s]$, and computes $U_{j}=h_{j}^{u_{j}}$.
It also chooses at random $w_{l} \in_{R} \mathbb{Z}_{p}$ and sets $c_{j}= m_{l,j} \cdot w_{l}  + u_{j} $, 
$C_{j}=h_{j}^{c_{j}}$ for $j \in [1,s]$, and $d= T_{m_{l}}^{w_{l}}$.
 Finally, it returns $\nu' = (U_{1},\cdots,U_{s},$ $C_{1},\cdots,C_{s},d,w_{l}) \in \mathbb{G}_{1}^{2s+1}$ to the TPA.
  
  For the operation:
\begin{enumerate}
\item \textit{Insertion:} $(l,m_{l},T_{m_{l}}) = (\frac{i_{1}+i_{2}}{2},m_{\frac{i_{1}+i_{2}}{2}},$

$T_{m_{\frac{i_{1}+i_{2}}{2}}})$ and $vnb_{l} = vnb_{\frac{i_{1}+i_{2}}{2}}=1$;
 \item \textit{Deletion:} $(l,m_{l},T_{m_{l}}) = (i,\_,\_)$ and $vnb_{l} = vnb_{i}=\_$, meaning that $m_{l}$, $T_{m_{l}}$ and $vnb_{l}$ are not required
  (the server uses $m_{i}$, $T_{m_{i}}$ and $vnb_{i}$ that are kept on its storage to generate $\nu'$);
\item \textit{Modification:} $(l,m_{l},T_{m_{l}}) =(i,m_{i}',T_{m_{i}'})$ and $ vnb_{l}=vnb_{i}' = vnb_{i}+1$.
\end{enumerate}

\noindent $\bullet$ $\textsf{CheckOp}(pk,\nu') \to 0/1$.
 The TPA has to check whether the following equation holds:
    \begin{eqnarray}
 e(d,g_{2}^{a}) \cdot e(\prod_{j=1}^{s} U_{j} ,g_{2}) &\stackrel{?}{=}& e ( H(l,vnb_{l})^{w_{l}},g_{2})     \cdot e(\prod_{j=1}^{s} C_{j},g_{2})
  \label{eqt15-iht}
  \end{eqnarray}
    If Eq. \ref{eqt15-iht} holds, then the TPA returns $1$ to the client; otherwise, it returns $0$ to the client.
   
 \noindent $\bullet$ $\textsf{GenProof}(pk,F,chal,\Sigma) \to \nu$.
 The TPA first chooses $I \subseteq (0,n+1) \cap \mathbb{Q}$, randomly chooses $|I|$ elements $v_{i} \in_{R} \mathbb{Z}_{p}$ and sets $chal=\{(i,v_{i})\}_{i \in I}$.
 
 After receiving $chal$, the server sets $F = \{m_{i}\}_{i \in I} \subset \mathbb{F}$ of blocks
 and $\Sigma = \{T_{m_{i}}\}_{i \in I} \subset \mathbb{E}$ which are the tags corresponding to the blocks in $F$.
 It then selects at random $r_{j} \in_{R} \mathbb{Z}_{p}$, for $j \in [1,s]$, and computes $R_{j}=h_{j}^{r_{j}}$.
 It also sets $b_{j}=\sum_{(i,v_{i}) \in chal} m_{i,j} \cdot v_{i}  + r_{j}$, $B_{j}=h_{j}^{b_{j}}$ for $j \in [1,s]$, 
 and $c =\prod_{(i,v_{i}) \in chal} T_{m_{i}}^{v_{i}}$.
 Finally, it returns $\nu = (R_{1},\cdots,R_{s},B_{1},\cdots,B_{s},c) \in \mathbb{G}_{1}^{2s+1}$ to the TPA.
  
 \noindent $\bullet$ $\textsf{CheckProof}(pk,chal,\nu) \to 0/1$.
  The TPA has to check whether the following equation holds:
  \begin{eqnarray}
    e(c,g_{2}^{a}) \cdot e(\prod_{j=1}^{s} R_{j} ,g_{2})  &\stackrel{?}{=}& e ( \prod_{\substack{(i,v_{i}) \\ \in chal}} H(i,vnb_{i})^{v_{i}},g_{2})   \cdot e(\prod_{j=1}^{s} B_{j},g_{2})
  \label{eqt25-iht}
  \end{eqnarray}
   If Eq. \ref{eqt25-iht} holds, then the TPA returns $1$ to the client; otherwise, it returns $0$ to the client.

 \paragraph{Correctness.}
Given the proof of data possession $\nu$ and the updating proof $\nu'$, we have:
 \begin{eqnarray*}
  e(c,g_{2}^{a}) \cdot e(\prod_{j=1}^{s} R_{j} ,g_{2})   &=&  e  ( \prod_{(i,v_{i})  \in chal} T_{m_{i}}^{v_{i}},g_{2}^{a} ) \cdot e(\prod_{j=1}^{s} h_{j}^{r_{j}} ,g_{2}) \\
  &=& e  ( \prod_{(i,v_{i}) \in chal} (H(i,vnb_{i}) \cdot \prod_{j=1}^{s} h_{j}^{m_{i,j} })^{-a \cdot v_{i}},g_{2}^{a}) \cdot e(\prod_{j=1}^{s} h_{j}^{r_{j}} ,g_{2}) \\
  &=& e  ( \prod_{(i,v_{i})  \in chal} H(i,vnb_{i})^{v_{i}},g_{2} ) \cdot e(\prod_{j=1}^{s} B_{j},g_{2}  ) 
 \end{eqnarray*}
 
 \begin{eqnarray*}
  e(d,g_{2}^{a})  \cdot e(\prod_{j=1}^{s} U_{j} ,g_{2})   &=&  e( T_{m_{l}}^{w_{l}},g_{2}^{a}) \cdot e(\prod_{j=1}^{s} h_{j}^{u_{j}} ,g_{2}) \\
  &=& e(H(l,vnb_{l}) \cdot \prod_{j=1}^{s} h_{j}^{m_{l,j} },g_{2}^{a})^{-a \cdot w_{l}}  \cdot e(\prod_{j=1}^{s} h_{j}^{u_{j}} ,g_{2}) \\
  &=&  e ( H(l,vnb_{l})^{w_{l}},g_{2}) \cdot e(\prod_{j=1}^{s} C_{j},g_{2})
 \end{eqnarray*}


\paragraph{N.B.}
The client or the TPA must store the values $vnb$ locally.
However, this does not incur more burden if we consider the values $vnb$ as bit strings.

\subsection{Security and Privacy Proofs}

\subsubsection{Security Proof against the Server}
\label{secproofiht}

\begin{theorem}
 Let $\mathcal{A}$ be a PPT adversary that has advantage $\epsilon$ against the IHT-based DPDP scheme with PV and DP.
 Suppose that $\mathcal{A}$ makes a total of $q_{H} >0$ queries to $H$.
 Then, there is a challenger $\mathcal{B}$ that solves the Computational Diffie-Hellman (CDH) and Discrete Logarithm (DL) problems with advantage $\epsilon' = \mathcal{O}(\epsilon)$.
\end{theorem}

\noindent For any PPT adversary $\mathcal{A}$ who wins the game, 
there is a challenger $\mathcal{B}$ that wants to break the CDH and DL problems by interacting with $\mathcal{A}$ as follows:

 \noindent $\diamond$
 \textit{KeyGen.} 
 $\mathcal{B}$ runs $\textsf{GroupGen}(\lambda) \to (p,\mathbb{G},\mathbb{G}_{T},e,g)$.
Then, it is given the CDH instance tuple $(g,g^{a},$ $g^{b})$ where $<g>=\mathbb{G}$, chooses two exponents $x,y \in \mathbb{Z}_{p}$ and computes $g_{1} = g^{x}$ and $g_{2} = g^{y}$.
It also sets $\mathbb{G}_{1} =<g_{1}>$ and $\mathbb{G}_{2} =<g_{2}>$.
Note that $(g^{a})^{x} = g_{1}^{a}$, $(g^{b})^{x} = g_{1}^{b}$, $(g^{a})^{y} = g_{2}^{a}$ and $(g^{b})^{y} = g_{2}^{b}$.
$\mathcal{B}$ chooses $\beta_{j},\gamma_{j} \in_{R} \mathbb{Z}_{p}$ and sets $h_{j} = g_{1}^{\beta_{j}} \cdot (g_{1}^{b})^{\gamma_{j}}$ for $j \in [1,s]$.
Let a hash function $H: \mathbb{Q} \times \mathbb{N} \to \mathbb{G}_{1}$ be controlled by $\mathcal{B}$ as follows.
Upon receiving a query $(i_{l'},vnb_{i_{l'}})$ to $H$ for some $l' \in [1,q_{H}]$:
\begin{enumerate}
\item If $((i_{l'},vnb_{i_{l'}}),\theta_{l'},W_{l'})$ exists in $L_{H}$, return $W_{l'}$; 
\item Otherwise, choose $\beta_{j},\gamma_{j} \in_{R} \mathbb{Z}_{p}$ and set $h_{j} = g_{1}^{\beta_{j}} \cdot (g_{1}^{b})^{\gamma_{j}}$ for $j \in [1,s]$.
  For each $i_{l'}$, choose $\theta_{l'} \in_{R} \mathbb{Z}_{p}$ at random and set
  \[ W_{l'}=\frac{g_{1}^{\theta_{l'}}}{g_{1}^{ \sum_{j=1}^{s} \beta_{j} m_{i_{l'},j}} (g_{1}^{b})^{ \sum_{j=1}^{s} \gamma_{j} m_{i_{l'},j} }} \]
  for a given block $m_{i_{l'}} = (m_{i_{l'},1} , \cdots, m_{i_{l'},s})$.
  Put $((i_{l'},vnb_{i_{l'}}),\theta_{l'},W_{l'})$ in $L_{H}$ and return $W_{l'}$.
\end{enumerate}

  $\mathcal{B}$ sets the public key $pk= (p,\mathbb{G}_{1},\mathbb{G}_{2},\mathbb{G}_{T},e,$ $g_{1},g_{2},h_{1},\cdots,h_{s},g_{2}^{a},H)$ and forwards it to $\mathcal{A}$.
$\mathcal{B}$ keeps $g_{1}^{a}$, $g_{1}^{b}$ and $g_{2}^{b}$ secret.

 \noindent $\diamond$
\textit{Adaptive Queries.}
$\mathcal{A}$ has first access to $\mathcal{O}_{TG}$ as follows. 
 It first adaptively selects blocks $m_{i}=(m_{i,1},\cdots,m_{i,s})$, for $i \in [1,n]$.
 Then, $\mathcal{B}$ computes $T_{m_{i}} = (W \cdot \prod_{j=1}^{s} h_{j}^{m_{i,j}})^{-sk} =  (W \cdot \prod_{j=1}^{s} $ $h_{j}^{m_{i,j}})^{-a}$, such that
 if $((i,vnb_{i}),\theta,W)$ exists in $L_{H}$, then $W$ is used to compute $T_{m_{i}}$. Otherwise, $\theta \in_{R} \mathbb{Z}_{p}$ is chosen at random, 
 $W=\frac{g_{1}^{\theta}}{g_{1}^{ \sum_{j=1}^{s} \beta_{j} m_{i,j}} (g_{1}^{b})^{ \sum_{j=1}^{s} \gamma_{j} m_{i,j} }}$ is computed for 
 $h_{j} =g_{1}^{\beta_{j}} \cdot (g_{1}^{b})^{\gamma_{j}}$, $((i,vnb_{i}),\theta,W)$ is put in $L_{H}$ and $W$ is used to compute $T_{m_{i}}$.
Note that we have
 \begin{eqnarray*}
 \prod_{j=1}^{s} h_{j}^{m_{i,j}} \cdot H(i,vnb_{i})  &=& (\prod_{j=1}^{s} h_{j}^{m_{i,j}}) \cdot \frac{g_{1}^{\theta}}{g_{1}^{ \sum_{j=1}^{s} \beta_{j} m_{i,j}} \cdot (g_{1}^{b})^{ \sum_{j=1}^{s} \gamma_{j} m_{i,j} }} \\
  &=& \frac{ g_{1}^{\sum_{j=1}^{s} \beta_{j} m_{i,j}} (g_{1}^{b})^{ \sum_{j=1}^{s} \gamma_{j} m_{i,j} } \cdot  g_{1}^{\theta} }{g_{1}^{ \sum_{j=1}^{s} \beta_{j} m_{i,j}} \cdot (g_{1}^{b})^{ \sum_{j=1}^{s} \gamma_{j} m_{i,j} }} = g_{1}^{\theta}
 \end{eqnarray*}
 and so, $T_{m_{i}} = (H(i,vnb_{i}) \cdot \prod_{j=1}^{s} h_{j}^{m_{i,j}} )^{-sk} = (H(i,vnb_{i}) \cdot \prod_{j=1}^{s} h_{j}^{m_{i,j}} )^{-a} = (g_{1}^{a})^{- \theta}$.

 $\mathcal{B}$ gives the blocks and tags to $\mathcal{A}$.
 The latter sets an ordered collection $\mathbb{F}=\{m_{i}\}_{i \in [1,n]}$ of blocks and an ordered collection 
 $\mathbb{E}=\{T_{m_{i}}$ $\}_{i \in [1,n]}$ which are the tags corresponding to the blocks in $\mathbb{F}$.
 
 $\mathcal{A}$ has also access to $\mathcal{O}_{DOP}$ as follows. 
 Repeatedly, $\mathcal{A}$ selects a block $m_{l}$ and the corresponding $info_{l}$ and forwards them to $\mathcal{B}$.
 Here, $l$ denotes the rank where $\mathcal{A}$ wants the data operation to be performed: $l$ is equal to $\frac{i_{1}+i_{2}}{2}$ for an insertion 
 and to $i$ for a deletion or a modification.
We recall that only the rank is needed for a deletion and the version number $vnb_{l}$ increases by $1$ for a modification.
 Then, $\mathcal{A}$ outputs two new ordered collections $\mathbb{F}'$ and $\mathbb{E}'$, and a corresponding updating proof 
 $\nu'=(U_{1},\cdots,U_{s},C_{1},\cdots,C_{s},d,w_{l})$, such that
 $w_{l} \in_{R} \mathbb{Z}_{p}$, $d  = T_{m_{l}}^{w_{l}}$, 
 and for $j \in [1,s]$, $u_{j} \in_{R} \mathbb{Z}_{p}$, $U_{j} =h_{j}^{u_{j}}$, 
 $c_{j}= m_{l,j} \cdot w_{l} + u_{j}$ and $C_{j}=h_{j}^{c_{j}}$.
 $\mathcal{B}$ runs $\textsf{CheckOp}$ on $\nu'$ and sends the answer to $\mathcal{A}$.
 If the answer is $0$, then $\mathcal{B}$ aborts; otherwise, it proceeds.
 
 
 \noindent $\diamond$
\textit{Challenge.} 
 $\mathcal{A}$ selects $m_{i}^{*}$ and $info_{i}^{*}$, for $i \in \mathcal{I} \subseteq (0,n+1) \cap \mathbb{Q}$, 
 and forwards them to $\mathcal{B}$ who checks the data operations.
 In particular, the first $info_{i}^{*}$ indicates a full re-write.
 
 $\mathcal{B}$ chooses a subset $I \subseteq \mathcal{I}$, randomly selects $|I|$ elements $v_{i} \in_{R} \mathbb{Z}_{p}$ and sets $chal=\{(i,v_{i})\}_{i \in I}$.
 It forwards $chal$ as a challenge to $\mathcal{A}$.
 
 \noindent $\diamond$
 \textit{Forgery.} 
 Upon receiving $chal$, the resulting proof of data possession on the correct stored file $m$ should be $\nu=(R_{1},\cdots,R_{s},B_{1},\cdots,B_{s},c)$ 
 and pass the Eq. \ref{eqt25-iht}. 
 However, $\mathcal{A}$ generates a proof of data possession on an incorrect stored file $\tilde{m}$ as $\tilde{\nu}=(\tilde{R}_{1},\cdots,\tilde{R}_{s},
 \tilde{B}_{1},\cdots,\tilde{B}_{s},\tilde{c})$, such that
 $\tilde{r}_{j} \in_{R} \mathbb{Z}_{p}$, $\tilde{R}_{j} =h_{j}^{\tilde{r}_{j}}$, 
 $\tilde{b}_{j}=\sum_{(i,v_{i}) \in chal}$ $ \tilde{m}_{i,j} \cdot v_{i} + \tilde{r}_{j}$ and $\tilde{B}_{j}=h_{j}^{\tilde{b}_{j}}$, for $j \in [1,s]$.
 It also sets $\tilde{c}  =\prod_{(i,v_{i}) \in chal} T_{\tilde{m}_{i}}^{v_{i}}$.
 Finally, it returns $\tilde{\nu}$ to $\mathcal{B}$.
 If $\tilde{\nu}$ still pass the verification, then $\mathcal{A}$ wins.
 Otherwise, it fails.

\paragraph{Analysis.}
We define $\Delta r_{j}=\tilde{r}_{j}-r_{j}$, $\Delta b_{j}=\tilde{b}_{j}-b_{j} =  \sum_{(i,v_{i}) \in chal} ( \tilde{m}_{i,j}- m_{i,j}) v_{i} +\Delta r_{j}$
 and $\Delta \mu_{j}$ $=\sum_{(i,v_{i}) \in chal} ( \tilde{m}_{i,j}- m_{i,j}) v_{i}$, for $j \in [1,s]$.
 Note that $r_{j}$ and $b_{j}$ are the elements of a honest proof of data possession $\nu$ such that $r_{j} \in_{R} \mathbb{Z}_{p}$ and
 $b_{j} = \sum_{(i,v_{i}) \in chal} m_{i,j} \cdot v_{i} + r_{j}$ where $m_{i,j}$ are the actual sectors (not the ones that $\mathcal{A}$ claims to have).
 
We prove that if $\mathcal{A}$ can win the game, then solutions to the CDH and DL problems are found, which contradicts the assumption that 
 the CDH and DL problems are hard in $\mathbb{G}$ and $\mathbb{G}_{1}$ respectively. 
 Let assume that $\mathcal{A}$ wins the game. 
 We recall that if $\mathcal{A}$ wins then $\mathcal{B}$ can extract the actual blocks $\{m_{i}\}_{(i,v_{i}) \in chal}$ in polynomially-many interactions with $\mathcal{A}$.
 Wlog, suppose that $chal = \{(i,v_{i})\}$, meaning the challenge contains only one block.
 \\
 
  \noindent $\circ$ \textit{First case ($\tilde{c} \neq c$):} According to Eq. \ref{eqt25-iht}, we have
 \begin{eqnarray*}
  e( \frac{ \tilde{c}}{c} ,g_{2}) &=& e \left ( \frac{T_{\tilde{m}_{i}}}{T_{m_{i}}} ,g_{2} \right )^{v_{i}} 
   = e ( \prod_{j=1}^{s} h_{j}^{ \Delta \mu_{j}} , g_{2}^{-a}) =e ( \prod_{j=1}^{s} (g_{1}^{\beta_{j}}  \cdot (g_{1}^{b})^{\gamma_{j}})^{\Delta \mu_{j}} ,g_{2}^{-a}) 
 \end{eqnarray*}
 and so, we get that
 \[ e( \frac{ \tilde{c}}{c} \cdot  (g_{1}^{a} )^{\sum_{j=1}^{s} \beta_{j} \Delta \mu_{j}} ,g_{2})  = e(g_{1}^{b},g_{2}^{-a})^{\sum_{j=1}^{s} \gamma_{j} \Delta \mu_{j}}  \]
 meaning that we have found the solution to the CDH problem, that is
 \[(g_{1}^{b})^{a} = (g^{x})^{ab}=  (\frac{ \tilde{c}}{c} \cdot  (g_{1}^{a} )^{\sum_{j=1}^{s} \beta_{j} \Delta \mu_{j}})^{ \frac{-1}{{\sum_{j=1}^{s} \gamma_{j} \Delta \mu_{j}}} }\]
 unless evaluating the exponent causes a divide-by-zero.
 Nevertheless, we notice that not all of the $\Delta \mu_{j}$ can be zero 
 (indeed, if $\mu_{j} = m_{i,j} v_{i} = \tilde{\mu}_{j} = \tilde{m}_{i,j} v_{i}$ for $j \in [1,s]$,
 then $c = \tilde{c}$ which contradicts
 the hypothesis), and the $\gamma_{j}$ are information theoretically hidden from $\mathcal{A}$ (Pedersen commitments), so the denominator is zero only with probability
 $1/p$, which is negligible.
 Finally, since $\mathcal{B}$ knows the exponent $x$ such that $g_{1}=g^{x}$, it can directly compute 
 \[ ((\frac{ \tilde{c}}{c} \cdot  (g_{1}^{a} )^{\sum_{j=1}^{s} \beta_{j} \Delta \mu_{j}})^{ \frac{-1}{{\sum_{j=1}^{s} \gamma_{j} \Delta \mu_{j}}} })^{ \frac{1}{x} } \]
 and obtains $g^{ab}$.
 Thus, if $\mathcal{A}$ wins the game, then a solution to the CDH problem can be found with probability equal to $1 - 1/p$.
 
 \noindent $\circ$ \textit{Second Case ($\tilde{c} = c$):} According to Eq. \ref{eqt25-iht}, we have
  $e(\tilde{c},g_{2}^{a}) =  e ( H(i,vnb_{i})^{v_{i}},g_{2}) \cdot  e(\prod_{j=1}^{s} \tilde{B}_{j},$ $g_{2}) \cdot e(\prod_{j=1}^{s} \tilde{R}_{j} ,g_{2})^{-1}$.
  Since the proof $\nu=(R_{1},$ $\cdots,R_{s},B_{1},\cdots,B_{s},c)$ is a correct one, we also have
$e(c,g_{2}^{a}) =  e ( H(i,vnb_{i})^{v_{i}},g_{2}) \cdot e(\prod_{j=1}^{s} B_{j},$ $g_{2}) \cdot e(\prod_{j=1}^{s} R_{j} ,g_{2})^{-1}$. 
  We recall that $chal = \{(i,v_{i})\}$.
From the previous analysis step, we know that $\tilde{c}=c$.
Therefore, we get that $\prod_{j=1}^{s} \tilde{B}_{j} \cdot (\prod_{j=1}^{s} \tilde{R}_{j} )^{-1}=  \prod_{j=1}^{s} B_{j} \cdot (\prod_{j=1}^{s} R_{j})^{-1}$. 
We can re-write as $\prod_{j=1}^{s} h_{j}^{\tilde{b}_{j} - \tilde{r}_{j}}=  \prod_{j=1}^{s} h_{j}^{b_{j} -r_{j}}$
    or even as $\prod_{j=1}^{s} h_{j}^{\Delta b_{j} - \Delta r_{j}}= \prod_{j=1}^{s} h_{j}^{\Delta \mu_{j}} =1$.
  For $g_{1}, h $ $\in \mathbb{G}_{1}$, there exists $\xi \in \mathbb{Z}_{p}$ such that $h=g_{1}^{\xi}$ since $\mathbb{G}_{1}$ is a cyclic group. 
Wlog, given $g_{1}, h \in \mathbb{G}_{1}$, each $h_{j}$ could randomly
and correctly be generated by computing $h_{j} = g_{1}^{y_{j}} \cdot h^{z_{j}} \in \mathbb{G}_{1}$
such that $y_{j}$ and $z_{j}$ are random values in $\mathbb{Z}_{p}$. Then, we have
$ 1= \prod_{j=1}^{s} h_{j}^{\Delta \mu_{j}} = \prod_{j=1}^{s} (g_{1}^{y_{j}} \cdot h^{z_{j}})^{\Delta \mu_{j}} = g_{1}^{\sum_{j=1}^{s} y_{j} \cdot \Delta \mu_{j}} \cdot
h^{\sum_{j=1}^{s} z_{j} \cdot \Delta \mu_{j} } $.
Clearly, we can find a solution to the DL problem. More specifically, given $g_{1},h=g_{1}^{\xi} \in \mathbb{G}_{1}$, we can compute
$ h = g_{1}^{\frac{\sum_{j=1}^{s} y_{j} \cdot \Delta \mu_{j}}{\sum_{j=1}^{s} z_{j} \cdot \Delta \mu_{j}}} = g_{1}^{\xi}$
unless the denominator is zero.
However, not all of the $\Delta \mu_{j}$ can be zero and the $z_{j}$ are information theoretically hidden from $\mathcal{A}$, so the denominator is only zero 
with probability $1/p$, which is negligible.
Thus, if $\mathcal{A}$ wins the game, then a solution to the DL problem can be found with probability equal to $1 - 1/p$.

Therefore, for $\mathcal{A}$, it is computationally infeasible to win the game and generate an incorrect proof of data possession which can pass the verification.
\\

The simulation of $\mathcal{O}_{TG}$ is perfect. 
The simulation of $\mathcal{O}_{DOP}$ is almost perfect unless $\mathcal{B}$ aborts. 
This happens when the data operation was not correctly performed. As previously, we can prove that if $\mathcal{A}$ can pass the updating proof,
then solutions to the CDH and DL problems are found.
Following the above analysis and according to Eq. \ref{eqt15-iht}, if $\mathcal{A}$ generates an incorrect updating proof which can pass the verification, 
then solutions to the CDH and DL problems can be found with probability equal to $1 - \frac{1}{p}$ respectively.
Therefore, for $\mathcal{A}$, it is computationally infeasible to generate an incorrect updating proof which can pass the verification.
The proof is completed.

\subsubsection{First Data Privacy Proof against the TPA}

\begin{theorem}
 Let $\mathcal{A}$ be a PPT adversary that has advantage $\epsilon$ against the IHT-based DPDP scheme with PV and DP.
 Suppose that $\mathcal{A}$ makes a total of $q_{H} >0$ queries to $H$.
 Then, there is a challenger $\mathcal{B}$ that solves the CDH problem with advantage $\epsilon' = \mathcal{O}(\epsilon)$.
\end{theorem}

\noindent 
 For any PPT adversary $\mathcal{A}$ who wins the game, there is a challenger $\mathcal{B}$ 
 that wants to break the CDH problem by interacting with $\mathcal{A}$ as follows:

 \noindent $\diamond$
 \textit{Setup.}
$\mathcal{B}$ runs $\textsf{GroupGen}(\lambda) \to (p,\mathbb{G},\mathbb{G}_{T},e,g)$.
Then, it is given the CDH instance tuple $(g,g^{a},$ $g^{b})$, chooses two exponents $x,y \in \mathbb{Z}_{p}$ and computes $g_{1} = g^{x}$ and $g_{2} = g^{y}$.
It also sets $\mathbb{G}_{1} =<g_{1}>$ and $\mathbb{G}_{2} =<g_{2}>$.
Note that $(g^{a})^{x} = g_{1}^{a}$, $(g^{b})^{x} = g_{1}^{b}$, $(g^{a})^{y} = g_{2}^{a}$ and $(g^{b})^{y} = g_{2}^{b}$.
$\mathcal{B}$ chooses $\beta_{j},\gamma_{j} \in_{R} \mathbb{Z}_{p}$ and sets $h_{j} = g_{1}^{\beta_{j}} \cdot (g_{1}^{b})^{\gamma_{j}}$ for $j \in [1,s]$.
Let a hash function $H: \mathbb{Q} \times \mathbb{N} \to \mathbb{G}_{1}$ be controlled by $\mathcal{B}$ as follows.
Upon receiving a query $(i_{l},vnb_{i_{l}})$ to the random oracle $H$ for some $l \in [1,q_{H}]$:
\begin{enumerate}
\item If $((i_{l},vnb_{i_{l}}),\theta_{l},W_{l})$ exists in $L_{H}$, return $W_{l}$;
\item Otherwise, choose $\beta_{j},\gamma_{j} \in_{R} \mathbb{Z}_{p}$ and set $h_{j} = g_{1}^{\beta_{j}} \cdot (g_{1}^{b})^{\gamma_{j}}$ for $j \in [1,s]$.
  For each $i_{l}$, choose $\theta_{l} \in_{R} \mathbb{Z}_{p}$ at random and set
  \[ W_{l}=\frac{g_{1}^{\theta_{l}}}{g_{1}^{ \sum_{j=1}^{s} \beta_{j} m_{i_{l},j}} (g_{1}^{b})^{ \sum_{j=1}^{s} \gamma_{j} m_{i_{l},j} }} \]
  for a given block $m_{i_{l}} = (m_{i_{l},1} , \cdots, m_{i_{l},s})$.
  Put $((i_{l},vnb_{i_{l}}),\theta_{l},W_{l})$ in $L_{H}$ and return $W_{l}$.
\end{enumerate}

  $\mathcal{B}$ sets the public key $pk= (p,\mathbb{G}_{1},\mathbb{G}_{2},\mathbb{G}_{T},e,$ $g_{1},g_{2},h_{1},\cdots,h_{s},g_{2}^{a},H)$ and forwards it to $\mathcal{A}$.
It keeps $g_{1}^{a}$, $g_{1}^{b}$ and $g_{2}^{b}$ secret.

 \noindent $\diamond$
 \textit{Queries.} $\mathcal{A}$ makes queries as follows. 
 It first adaptively selects blocks $m_{i}=(m_{i,1},\cdots,m_{i,s})$, for $i \in [1,n]$.
 Then, $\mathcal{B}$ computes $T_{m_{i}} = (W \cdot \prod_{j=1}^{s} h_{j}^{m_{i,j}})^{-sk} =  (W \cdot \prod_{j=1}^{s} h_{j}^{m_{i,j}})^{-a}$, such that
 if $((i,vnb_{i}),\theta,W)$ exists in $L_{H}$, then $W$ is used to compute $T_{m_{i}}$. Otherwise, $\theta \in_{R} \mathbb{Z}_{p}$ is chosen at random, 
 $W=\frac{g_{1}^{\theta}}{g_{1}^{ \sum_{j=1}^{s} \beta_{j} m_{i,j}} (g_{1}^{b})^{ \sum_{j=1}^{s} \gamma_{j} m_{i,j} }}$ is computed for
 $h_{j} =g_{1}^{\beta_{j}} \cdot (g_{1}^{b})^{\gamma_{j}}$, $((i,vnb_{i}),\theta,W)$ is put in $L_{H}$ and $W$ is used to compute $T_{m_{i}}$.
Note that we have
 \begin{eqnarray*}
  \prod_{j=1}^{s} h_{j}^{m_{i,j}} \cdot H(i,vnb_{i}) &=& (\prod_{j=1}^{s} h_{j}^{m_{i,j}}) \cdot \frac{g_{1}^{\theta}}{g_{1}^{ \sum_{j=1}^{s} \beta_{j} m_{i,j}} \cdot (g_{1}^{b})^{ \sum_{j=1}^{s} \gamma_{j} m_{i,j} }} \\
  &=& \frac{ g_{1}^{\sum_{j=1}^{s} \beta_{j} m_{i,j}} (g_{1}^{b})^{ \sum_{j=1}^{s} \gamma_{j} m_{i,j} } \cdot  g_{1}^{\theta} }{g_{1}^{ \sum_{j=1}^{s} \beta_{j} m_{i,j}}
  \cdot (g_{1}^{b})^{ \sum_{j=1}^{s} \gamma_{j} m_{i,j} }} = g_{1}^{\theta}
 \end{eqnarray*}
 and so, $T_{m_{i}} = (H(i,vnb_{i}) \cdot \prod_{j=1}^{s} h_{j}^{m_{i,j}} )^{-sk} = (H(i,vnb_{i}) \cdot \prod_{j=1}^{s} h_{j}^{m_{i,j}} )^{-a} = (g_{1}^{a})^{- \theta}$.

 $\mathcal{B}$ gives the blocks and tags to $\mathcal{A}$ and the latter sets two ordered collections $\mathbb{F}=\{m_{i}$ $\}_{i \in [1,n]}$ and
 $\mathbb{E}=\{T_{m_{i}}\}_{i \in [1,n]}$.

\noindent $\diamond$
\textit{Challenge.} $\mathcal{A}$ submits a challenge $chal = \{(i,$ $v_{i})\}_{i \in I}$.
Wlog, we suppose there is only $i$ in $I$ and we write $chal = \{(i,v_{i})\}$.
$\mathcal{A}$ gives an ordered collection $F = \{\tilde{m}_{i}\} \cap \mathbb{F} = \emptyset$ of the blocks determined by $chal$, 
and an ordered collection $\Sigma =\{T_{\tilde{m}_{i}}\}  \cap \mathbb{E} = \emptyset$
of the corresponding tags. Note that there are only $\tilde{m}_{i}$ in $F$ and $T_{\tilde{m}_{i}}$ in $\Sigma$.
 
  \noindent $\diamond$
 \textit{Generation of the Proof.} Upon receiving $chal $ $= \{(i,v_{i})\}$, $F =\{\tilde{m}_{i}\} $ and $\Sigma = \{T_{\tilde{m}_{i}}\}$,
 $\mathcal{B}$ generates the proof of data possession $\tilde{\nu} =(\tilde{R}_{1},\cdots,\tilde{R}_{s},$ $
 \tilde{B}_{1},\cdots,\tilde{B}_{s},\tilde{c})$, such that
 $\tilde{r}_{j} \in_{R} \mathbb{Z}_{p}$, $\tilde{R}_{j} =h_{j}^{\tilde{r}_{j}}$, 
 $\tilde{b}_{j}=\sum_{(i,v_{i}) \in chal} \tilde{m}_{i,j} \cdot v_{i} + \tilde{r}_{j}$ and $\tilde{B}_{j}=h_{j}^{\tilde{b}_{j}}$, for $j \in [1,s]$.
 It also sets $\tilde{c}  =\prod_{(i,v_{i}) \in chal} T_{\tilde{m}_{i}}^{v_{i}}$.
 
 Finally, $\mathcal{B}$ runs {\sf CheckProof} on $\tilde{\nu}$.
 If $\tilde{\nu}$ still pass the verification, then $\mathcal{A}$ wins.
 Otherwise, it fails.
 
 \paragraph{Analysis.}
Given a honest file $m_{i}$ and the corresponding tags $T_{m_{i}}$, let $\nu = (R_{1},\cdots,R_{s},B_{1},\cdots,$ $B_{s},c)$
be a honest proof of data possession that pass the verification.

We define $\Delta r_{j}=\tilde{r}_{j}-r_{j}$,
$\Delta b_{j}=\tilde{b}_{j}-b_{j} =  \sum_{(i,v_{i}) \in chal} ( \tilde{m}_{i,j}- m_{i,j}) v_{i} +\Delta r_{j} =  ( \tilde{m}_{i,j}- m_{i,j}) $ $v_{i} +\Delta r_{j}$
 and $\Delta \mu_{j}=\sum_{(i,v_{i}) \in chal} ( \tilde{m}_{i,j}- m_{i,j}) v_{i} $ $=(\tilde{m}_{i,j}- m_{i,j}) v_{i}$, for $j \in [1,s]$.
 Note that $r_{j}$ and $b_{j}$ are the elements of a honest proof of data possession $\nu$ such that $r_{j} \in_{R} \mathbb{Z}_{p}$ and
 $b_{j} = \sum_{(i,v_{i}) \in chal} m_{i,j} \cdot v_{i} + r_{j} =m_{i,j} \cdot v_{i} + r_{j}$ where $m_{i,j}$ are the actual sectors 
 (not the ones that $\mathcal{A}$ claims to have).
 
We prove that if $\mathcal{A}$ can win the game, then a solution to the CDH problem is found, which contradicts the assumption that 
 the CDH problem is hard in $\mathbb{G}$. 
 Let assume that $\mathcal{A}$ wins the game.
 
Since $T_{\tilde{m}_{i}} \neq T_{m_{i}}$, and so $\tilde{c} \neq c$, We have
 \begin{eqnarray*}
  e( \frac{ \tilde{c}}{c} ,g_{2}) &=& e  ( \frac{T_{\tilde{m}_{i}}}{T_{m_{i}}} ,g_{2} )^{v_{i}} 
   = e ( \prod_{j=1}^{s} h_{j}^{ \Delta \mu_{j}} , g_{2}^{-a}) = e ( \prod_{j=1}^{s} (g_{1}^{\beta_{j}}  \cdot (g_{1}^{b})^{\gamma_{j}})^{\Delta \mu_{j}} ,g_{2}^{-a}) 
 \end{eqnarray*}
 and so, we get
 \[ e( \frac{ \tilde{c}}{c} \cdot  (g_{1}^{a} )^{\sum_{j=1}^{s} \beta_{j} \Delta \mu_{j}} ,g_{2})  = e(g_{1}^{b},g_{2}^{-a})^{\sum_{j=1}^{s} \gamma_{j} \Delta \mu_{j}}  \]
 meaning that we have found the solution to the CDH problem, that is
 \[(g_{1}^{b})^{a} = (g^{x})^{ab}=  (\frac{ \tilde{c}}{c} \cdot  (g_{1}^{a} )^{\sum_{j=1}^{s} \beta_{j} \Delta \mu_{j}})^{ \frac{-1}{{\sum_{j=1}^{s} \gamma_{j} \Delta \mu_{j}}} }\]
 unless evaluating the exponent causes a divide-by-zero.
 Nevertheless, we notice that not all of the $\Delta \mu_{j}$ can be zero (indeed, if $\mu_{j} = \tilde{\mu}_{j}$ for each $j \in [1,s]$, 
 then $c = \tilde{c}$ which contradicts
 the hypothesis), and the $\gamma_{j}$ are information theoretically hidden from $\mathcal{A}$ (Pedersen commitments), so the denominator is zero only with probability
 $1/p$, which is negligible.
 Finally, since $\mathcal{B}$ knows the exponent $x$ such that $g_{1}=g^{x}$, it can directly compute 
 \[((\frac{ \tilde{c}}{c} \cdot  (g_{1}^{a} )^{\sum_{j=1}^{s} \beta_{j} \Delta \mu_{j}})^{ \frac{-1}{{\sum_{j=1}^{s} \gamma_{j} \Delta \mu_{j}}} })^{ \frac{1}{x} } \]
 and obtains $g^{ab}$.
 Thus, if $\mathcal{A}$ wins the game, then a solution to the CDH problem can be found with probability equal to $1 - 1/p$.


\subsection{Performance}

\noindent We compare the IHT-based scheme with the original scheme proposed in \cite{GSP15}.
First, the client and TPA obviously have to store more information by keeping the IHT.
Nevertheless, we stress that in any case, the client and TPA should maintain a rank list. Indeed, they need some information about the stored data in order to select some data blocks to be challenged.
We recall that the challenge consists of pairs of the form ``(rank, random element)''.
By appending an integer and sometimes an auxiliary comment (only in case of deletions) to each rank, the extra burden is not excessive.
Therefore, such table does slightly affect the client's as well as TPA's local storages. The communication between the client and TPA rather increases since the client should send more elements to the TPA
in order to keep the table updated.

Second, the client has to perform extra computation when generating the verification metadata: for each file block $m_{i}$, he/she has to compute $H(i,vnb_{i})$.
However, the communication between the client and server overhead does not increase.

Third, the TPA needs to compute an extra pairing $e(H(i,vnb_{i}),g_{2})^{w_{i}}$ in order to check that the server correctly performed a data operation requested by the client.
The TPA also has to compute $|I|$ multiplications in $\mathbb{G}_{1}$ and one extra pairing when checking the proof of data possession: for each challenge $chal = \{(i,$ $v_{i})\}_{i \in I}$, it calculates $\prod_{(i,v{i}) \in chal}
H(i,vnb_{i})$ as well as the pairing $e(\prod_{(i,v{i}) \in chal} H(i,vnb_{i})^{v_{i}},$ $g_{2})$. This gives a constant total of four pairings in order to verify the data integrity instead of three, that is not a big loss in term of efficiency
and practicality.

Finally, apart the storage of a light table and computation of an extra pairing by the TPA for the verification of both the updating proof and proof of data possession,
the new construction for the DPDP scheme with PV and DP is still practical
by adopting asymmetric pairings to gain efficiency and by still reducing the group exponentiation and pairing operations.
In addition, this scheme still allows the TPA on behalf of the client to request the server for a proof of data possession on as many data blocks as 
possible at no extra cost, as in the scheme given in \cite{GSP15}.

\section{MHT-based DPDP scheme with PV and DP}
\label{mhtscheme}

\noindent A second solution to avoid the three attacks
is to implement a MHT \cite{M79} for each file. 
In a MHT, each internal node has always two children. 
For a leaf node $nd_{i}$ based on the block $m_{i}$, the assigned value is $H'(m_{i})$,
where the hash function $H' : \{0,1\}^{*} \to \mathbb{G}_{1}$ is seen as a random oracle. 
Note that the hash values are affected to the leaf nodes in the increasing order of the blocks:
$nd_{i}$ and $nd_{i+1}$ correspond to the hash of the blocks $m_{i}$ and $m_{i+1}$ respectively.
A parent node of $nd_{i}$ and $nd_{i+1}$ has a value computed as $H'(H'(m_{i}) || H'(m_{i+1}))$, where $||$ is the concatenation sign (for an odd rank $i$).
The Auxiliary Authentication Information (AAI) $\Omega_{i}$ of a leaf node $nd_{i}$ for $m_{i}$ is a set of hash values chosen from its upper levels, 
so that the root $rt$ can be computed using $(m_{i},\Omega_{i})$.

\subsection{MHT-based Construction}

\noindent Let $\textsf{DPDP}$ 
be a DPDP construction with PV and DP such as defined in Sec. \ref{schemedef} and \cite{GSP15}. 
Let $\textsf{SS}=(\textsf{Gen},\textsf{Sign},\textsf{Verify})$ be a strongly unforgeable digital signature scheme.
The MHT-based DPDP scheme with PV and DP construction is as follows:

 \noindent $\bullet$ $\textsf{MHT.KeyGen}(\lambda) \to ({\sf pk},{\sf sk})$.
 Let $\textsf{GroupGen}(\lambda) $ $\to (p,\mathbb{G}_{1},\mathbb{G}_{2},\mathbb{G}_{T},e,g_{1},g_{2})$ be run as follows.
On input the security parameter $\lambda$, \textsf{GroupGen}
generates the cyclic groups $\mathbb{G}_{1}$, $\mathbb{G}_{2}$ and $\mathbb{G}_{T}$ of prime order $p = p(\lambda)$ with the
bilinear map $e : \mathbb{G}_{1} \times \mathbb{G}_{2} \to \mathbb{G}_{T}$. 
Let $<g_{1}>=\mathbb{G}_{1}$ and $<g_{2}>=\mathbb{G}_{2}$.
 The client runs $\textsf{Gen}(\lambda) \to (pk_{\textsf{SS}},sk_{\textsf{SS}})$ and $\textsf{KeyGen}(\lambda) \to (pk,sk) =((p,\mathbb{G}_{1},\mathbb{G}_{2},\mathbb{G}_{T},e,g_{1},g_{2},$ $h_{1},\cdots,h_{s},g_{2}^{a}),a)$, where $h_{1},\cdots,h_{s} \in_{R} \mathbb{G}_{1}$ and $a \in_{R} \mathbb{Z}_{p}$ are randomly chosen.
 The client sets his/her public key ${\sf pk} = (pk,pk_{\textsf{SS}}) $
 and his/her secret key ${\sf sk} =(sk,sk_{\textsf{SS}}) $. 
 
\noindent $\bullet$  $\textsf{MHT.TagGen}({\sf pk},{\sf sk},m_{i}) \to T_{m_{i}}$.
 The client runs $n$ times $\textsf{TagGen}(pk,sk,m_{i}) \to T_{m_{i}}' = (\prod_{j=1}^{s} $ $ h_{j}^{m_{i,j}})^{-sk} =  (\prod_{j=1}^{s} h_{j}^{m_{i,j}})^{-a}$ for $i \in [1,n]$
 and obtains $T_{m}'= (T_{m_{1}}',\cdots,T_{m_{n}}') \in \mathbb{G}_{1}^{n}$.
  He/she also chooses a hash function $H' : \{0,1\}^{*} \to \mathbb{G}_{1}$ seen as a random oracle.
 Then, he/she creates the MHT regarding the file $m =(m_{1},\cdots,m_{n})$ as follows. He/she computes $H'(m_{i})$ and assigns it to the $i$-th leaf for $i \in [1,n]$.
He/she starts to construct the resulting MHT, and obtains the root $rt$.
 Finally, the client runs $\textsf{Sign}(sk_{\textsf{SS}},rt) \to \sigma_{rt}$.
 Using the hash values, he/she computes the tags as
 $T_{m_{i}} = H'(m_{i})^{-sk} \cdot T_{m_{i}}'= H'(m_{i})^{-a} \cdot \prod_{j=1}^{s} h_{j}^{- a \cdot m_{i,j}}$ for $i \in [1,n]$.

 Then, the client stores all the blocks $m_{i}$ in an ordered collection $\mathbb{F}$ and the corresponding tags $T_{m_{i}}$ 
 in an ordered collection $\mathbb{E}$. He/she forwards these two collections and $(H',\sigma_{rt})$ to the server.
 Once the server receives $(\mathbb{F},\mathbb{E},H')$, it generates the MHT. 
 It sends the resulting root $rt_{server}$ to the client.
 Upon getting the root $rt_{server}$, the client runs $\textsf{Verify}(pk_{\textsf{SS}},\sigma_{rt},$ $rt_{server}) \to 0/1$.
 If $0$, then the client aborts.
 Otherwise, he/she proceeds, deletes $(\mathbb{F},\mathbb{E},\sigma_{rt})$ from his/her local storage and keeps $H'$ for further data operations.
 
 \noindent $\bullet$  $\textsf{MHT.PerfOp}({\sf pk},\mathbb{F},\mathbb{E},R=(\mbox{operation},i),info$ $=(m_{i},T_{m_{i}},\sigma_{rt'})) \to (\mathbb{F'},\mathbb{E}',rt_{server}')$.
 First, the client sends a request $R=(\mbox{operation},i)$ to the server,
 that contains the type and rank of the operation. 
 Upon receiving $R$, the server selects the AAI $\Omega_{i}$ that the client needs in order to generate the root $rt'$ of the updated MHT, and sends it to the client.
 Once the client receives $\Omega_{i}$, 
 he/she first constructs the updated MHT. 
 He/she calculates the new root $rt'$ and runs $\textsf{Sign}(sk_{\textsf{SS}},rt') \to \sigma_{rt'}$.
 Then, the client sends $info=(m_{i},T_{m_{i}},\sigma_{rt'})$ (note that $m_{i}$ and $T_{m_{i}}$ are not needed for a deletion).
 After receiving $info$ from the client, the server first updates the MHT, calculates the new root
 $rt_{server}'$ and sends it to the client. 
 Upon getting the root $rt_{server}'$, the client runs $\textsf{Verify}(pk_{\textsf{SS}},\sigma_{rt'},rt_{server}') $ $\to 0/1$ .
 If $0$, then the client aborts. Otherwise, he/she proceeds and deletes $(m_{i},T_{m_{i}},\sigma_{rt'})$ from his/her local storage.
 
  
  For the operation:
\begin{enumerate}
\item \textit{Insertion:} $m_{i_{0}}$ is added before $m_{i}$ by placing $m_{i_{0}}$ at the $i$-th leaf node, and all the blocks from $m_{i}$
  are shifted to leaf nodes by $1$ to the right;
\item \textit{Deletion:} $m_{i}$ is removed from the $i$-th leaf node and all the blocks from $m_{i+1}$ are shifted to leaf nodes by $1$ to the left;
 \item \textit{Modification:} $m_{i}'$ simply replaces $m_{i}$ at the $i$-th leaf node.
  \end{enumerate}

 \noindent $\bullet$ $\textsf{MHT.GenProof}({\sf pk},F,chal,\Sigma) \to (\nu,rt_{server},$ $\{H'(m_{i}),\Omega_{i}\}_{i \in I})$.
  The TPA chooses a subset $I \subseteq [1,n_{max}]$ ($n_{max}$ is the maximum number of blocks after operations), 
  randomly chooses $|I|$ elements $v_{i} \in_{R} \mathbb{Z}_{p}$ and sets the challenge $chal=\{(i,v_{i})\}_{i \in I}$.
   Then, after receiving $chal$ and given $F = \{m_{i}\}_{i \in I} \subset \mathbb{F}$ and $\Sigma = \{T_{m_{i}}\}_{i \in I} $ $\subset \mathbb{E}$, 
   the server runs $\textsf{GenProof}(pk,F,chal,\Sigma) \to \nu$ such that $\nu = (R_{1},\cdots,R_{s},B_{1},\cdots,B_{s},c) $ $\in \mathbb{G}_{1}^{2s+1}$,
 where $r_{j} \in_{R} \mathbb{Z}_{p}$, $R_{j}=h_{1}^{r_{j}}$, $b_{j}=$

\noindent $\sum_{(i,v_{i}) \in chal} m_{i,j} \cdot v_{i}  + r_{j} \in \mathbb{Z}_{p}$ 
 and $B_{j}=h_{j}^{b_{j}}$ for $j \in [1,s]$, and $c =\prod_{(i,v_{i}) \in chal} T_{m_{i}}^{v_{i}}$.
   Moreover, the server prepares the latest version of the stored root's signature $\sigma_{rt}$ provided by the client, 
   the root $rt_{server}$ of the current MHT, the $H'(m_{i})$ and AAI $\Omega_{i}$ for the challenged blocks,
 such that the current MHT has been constructed using $\{H'(m_{i}),\Omega_{i}\}_{i \in I}$. 
 Finally, it returns $(\nu,\sigma_{rt},$ $rt_{server},\{H'(m_{i}),\Omega_{i}\}_{i \in I})$ to the TPA.
  
 \noindent $\bullet$ $\textsf{MHT.CheckProof}({\sf pk},chal,\nu,\sigma_{rt},rt_{server},\{H'$ $(m_{i}),\Omega_{i}\}_{i \in I}) \to 0/1$.  
  After receiving $\{H'(m_{i}),$ $\Omega_{i}\}_{i \in I}$ from the server, the TPA first constructs the MHT and calculates the root $rt_{TPA}$.
  It then checks that $rt_{server}=rt_{TPA}$. If not, then it aborts; 
  otherwise, it runs $\textsf{Verify}(pk_{\textsf{SS}},$ $\sigma_{rt},rt_{server}) \to 0/1$. If $0$, then the TPA aborts.
  Otherwise, it proceeds and checks whether the following equation holds:
 \begin{eqnarray}
  e(c,g_{2}^{a}) \cdot e(\prod_{j=1}^{s} R_{j} ,g_{2}) &\stackrel{?}{=}&  e(\prod_{(i,v_{i}) \in chal} H'(m_{i})^{v_{i}} ,g_{2})  \cdot e(\prod_{j=1}^{s} B_{j},g_{2})
 \label{eqt2-mht}
 \end{eqnarray}
  If Eq. \ref{eqt2-mht} holds, then the TPA returns $1$ to the client; otherwise, it returns $0$ to the client.

 \paragraph{Correctness.}
We suppose that the correctness holds for $\textsf{DPDP}$ and $\textsf{SS}$ protocols.
Given the proof of data possession $\nu$, we have:
\begin{eqnarray*}
  e(c,g_{2}^{a}) \cdot e(\prod_{j=1}^{s} R_{j} ,g_{2}) &=&  e  ( \prod_{(i,v_{i}) \in chal} T_{m_{i}}^{v_{i}},g_{2}^{a}  ) \cdot e(\prod_{j=1}^{s} h_{j}^{r_{j}} ,g_{2}) \\
 &=& e  ( \prod_{(i,v_{i}) \in chal} (H'(m_{i}) \cdot \prod_{j=1}^{s} h_{j}^{m_{i,j}})^{-a \cdot v_{i}},g_{2}^{a} )  \cdot e(\prod_{j=1}^{s} h_{j}^{r_{j}} ,g_{2}) \\
 &=& e  ( \prod_{(i,v_{i}) \in chal} H'(m_{i})^{v_{i}},g_{2}  ) \cdot e(\prod_{j=1}^{s} B_{j},g_{2}) 
\end{eqnarray*}

\paragraph{N.B.}
In \textsf{MHT.GenProof}, since $I$ is a subset of ranks, the server has to be given the appropriate $\{\Omega_{i}\}_{i \in I}$
 along with $\{H'(m_{i})\}_{i \in I}$ to obtain the current MHT and thus complete the proof generation.
Otherwise, the TPA won't get the proper MHT.

\subsection{Security and Privacy Proofs}

\subsubsection{Security Proof against the Server}

\begin{theorem}
 Let $\mathcal{A}$ be a PPT adversary that has advantage $\epsilon$ against the MHT-based DPDP scheme with PV and DP.
 Suppose that $\mathcal{A}$ makes a total of $q_{H'} >0$ queries to $H'$.
 Then, there is a challenger $\mathcal{B}$ that solves the CDH and DL problems with advantage $\epsilon' = \mathcal{O}(\epsilon)$.
\end{theorem}

\noindent For any PPT adversary $\mathcal{A}$ who wins the game, there is a challenger $\mathcal{B}$ that wants to break the CDH and DL problems
by interacting with $\mathcal{A}$ as follows:

 \noindent $\diamond$
 \textit{KeyGen.} 
This phase is similar to the one of the proof in Sec. \ref{secproofiht}, except that $H': \{0,1\}^{*} \to \mathbb{G}_{1}$ is controlled by $\mathcal{B}$ as follows.
 Upon receiving a query $m_{i_{l}}$ to the random oracle $H'$ for some $l \in [1,q_{H'}]$:
\begin{enumerate}
\item If $(m_{i_{l}},\theta_{l},W_{l})$ exists in $L_{H'}$, return $W_{l}$;
\item Otherwise, choose $\beta_{j},\gamma_{j} \in_{R} \mathbb{Z}_{p}$ and set $h_{j} = g_{1}^{\beta_{j}} \cdot (g_{1}^{b})^{\gamma_{j}}$ for $j \in [1,s]$.
  For each $i_{l}$, choose $\theta_{l} \in_{R} \mathbb{Z}_{p}$ at random and set
  \[W_{l}=\frac{g_{1}^{\theta_{l}}}{g_{1}^{ \sum_{j=1}^{s} \beta_{j} m_{i_{l},j}} (g_{1}^{b})^{ \sum_{j=1}^{s} \gamma_{j} m_{i_{l},j} }}\]
  for a given block $m_{i_{l}} = (m_{i_{l},1} , \cdots, m_{i_{l},s})$.
  Put $(m_{i_{l}},\theta_{l},W_{l})$ in $L_{H'}$ and return $W_{l}$.
\end{enumerate}

The hash function $H'$ and digital signature scheme \textsf{SS} are supposed to be collision resistant and strongly unforgeable respectively.
$\mathcal{B}$ gives $\mathcal{A}$ the public key ${\sf pk}$ that contains 
$pk = (p,\mathbb{G}_{1},\mathbb{G}_{2},\mathbb{G}_{T},e,g_{1},$ $g_{2},h_{1},\cdots,h_{s},g_{2}^{a})$ and $pk_{\textsf{SS}}$ $ \gets \textsf{Gen}(\lambda)$. 
$\mathcal{B}$ keeps $g_{1}^{a}$, $g_{1}^{b}$, $g_{2}^{b}$, $sk_{\textsf{SS}} \gets \textsf{Gen}(\lambda)$ and $H'$ secret.
 
\noindent $\diamond$
\textit{Adaptive Queries.}
 This phase is similar to the one of the proof in Sec. \ref{secproofiht}, except the following.
During the calls to $\mathcal{O}_{TG}$, $\mathcal{B}$ generates the tags and then 
creates the MHT resulting from $m_{i}$ and $rt$.
 It signs $rt$ by running $\sigma_{rt} \gets \textsf{Sign}(sk_{\textsf{SS}},rt)$.
  It finally gives the tags $T_{m_{i}}$, their corresponding $W$ resulting from calling $H'$ and $\sigma_{rt}$ to $\mathcal{A}$.
  
  During the calls to $\mathcal{O}_{DOP}$, $\mathcal{A}$ repeatedly selects $m_{i}$ 
  and $(R_{i},info_{i})$, and forwards them to $\mathcal{B}$. 
  The signature $\sigma_{rt} \gets \textsf{Sign}(sk_{\textsf{SS}},rt')$ of the root $rt'$ is included in $info_{i}$.
 Here, $i$ denotes the rank where $\mathcal{A}$ wants the operation to be performed.
 Then, $\mathcal{A}$ outputs two new ordered collections $\mathbb{F}'$ and $\mathbb{E}'$, and a new root $rt_{\mathcal{A}}'$ corresponding to the updated MHT.
  $\mathcal{B}$ runs $\textsf{Verify}$ on $\sigma_{rt'}$ and $rt_{\mathcal{A}}'$ and aborts if the answer is equal to $0$; it proceeds otherwise.
 
 \noindent $\diamond$
 \textit{Challenge.} 
 This phase is identical to the one of the security proof in Sec. \ref{secproofiht}, except that $info_{i}^{*}$, $R_{i}^{*}$ 
 and $i \in \mathcal{I} = [1,n']$, for $n' \ge n$, are given.
 
\noindent $\diamond$
 \textit{Forgery.} 
This phase is identical to the one of the security proof in Sec. \ref{secproofiht} except that we refer to Eq. \ref{eqt2-mht}.

\paragraph{Analysis.}
The first two parts of the analysis are identical to the ones of the security proof in Sec. \ref{secproofiht}.
 The last part slightly changes as follows.
The simulations of $\mathcal{O}_{TG}$ and $\mathcal{O}_{DOP}$ are perfect.
The proof is completed.

\subsubsection{Second Data Privacy Proof against the TPA}

\begin{theorem}
 Let $\mathcal{A}$ be a PPT adversary that has advantage $\epsilon$ against the MHT-based DPDP scheme with PV and DP.
  Suppose that $\mathcal{A}$ makes a total of $q_{H'} >0$ queries to $H'$.
 Then, there is a challenger $\mathcal{B}$ that solves the $(s+1)$-DDHE problem with advantage $\epsilon' = \mathcal{O}(\epsilon)$.
\end{theorem}

\noindent We presume that the digital signature scheme \textsf{SS} is strongly unforgeable and the hash function $H'$ is collision resistant. 
For any PPT adversary $\mathcal{A}$ who wins the game, there is a challenger $\mathcal{B}$ that wants to break the 
$(s+1)$-DDHE problem by interacting with $\mathcal{A}$ as follows:

\noindent $\diamond$
\textit{Setup.}
  $\mathcal{B}$ runs $\textsf{GroupGen}(\lambda) \to (p,\mathbb{G}_{1},\mathbb{G}_{2},\mathbb{G}_{T},$ $e,g_{1},g_{2})$ and receives the $(s+1)$-DDHE instance
 $(g_{1},g_{1}^{\beta},\cdots,g_{1}^{\beta^{s+1}},g_{2},g_{2}^{\beta},Z)$
where $<g_{1}>=\mathbb{G}_{1}$ and $<g_{2}>=\mathbb{G}_{2}$. 
 $\mathcal{B}$ sets $\mu=0$ when $Z=g_{1}^{\beta^{s+2}}$; 
 otherwise, it sets $\mu=1$ when $Z \in_{R} \mathbb{G}_{1}$.
Then, it randomly chooses $\xi_{1},\cdots,\xi_{s},\xi_{s+1} \in_{R} \mathbb{Z}_{p}$ and sets $h_{j}=(g_{1}^{\beta^{j}})^{\xi_{j}}$ for $j \in [1,s+1]$.
$\mathcal{B}$ also controls $H': \{0,1\}^{*} \to \mathbb{G}_{1}$ as follows.
Upon receiving a query $m_{i_{l}}$ to the random oracle $H'$ for some $l \in [1,q_{H'}]$:
\begin{enumerate}
\item If $(m_{i_{l}},\theta_{l},V_{l},W_{l})$ exists in $L_{H'}$, return $V_{l}$ and $W_{l}$;
\item Otherwise, choose $\theta_{l} \in_{R} \mathbb{Z}_{p}$ at random and compute $V_{l}=g_{1}^{- \theta_{l}}$ 
  and $W_{l}=h_{1}^{- \theta_{l} / \xi_{1}} =$

$ g_{1}^{-\beta \xi_{1} \theta_{l} /\xi_{1}}=g_{1}^{-\beta \theta_{l}}$.
  Put $(m_{i_{l}},\theta_{l},V_{l},W_{l})$ in $L_{H'}$ and return $W_{l}$.
\end{enumerate}
It sets the public key $pk= (p,\mathbb{G}_{1},\mathbb{G}_{2},\mathbb{G}_{T},e,$ $g_{1},g_{2},h_{1},\cdots,h_{s},g_{2}^{\beta})$.
$\mathcal{B}$ has also access to $\textsf{SS}$ and runs $\textsf{Gen}(\lambda)$ to obtain $(pk_{\textsf{SS}},sk_{\textsf{SS}})$.
$\mathcal{B}$ sets the public key ${\sf pk}=(pk,pk_{\textsf{SS}})$ and forwards it to $\mathcal{A}$.
$\mathcal{B}$ keeps $sk_{\textsf{SS}}$ and $H'$ secret.
The secret $sk$ is implicitly set as equal to $\beta$.

\noindent $\diamond$
\textit{Queries.}
 $\mathcal{A}$ makes queries as follows.
 $\mathcal{A}$ first selects a file $m =(m_{1},\cdots,m_{n})$ and sends it to $\mathcal{B}$.
 Then, $\mathcal{B}$ splits each block $m_{i}$ into $s$ sectors $m_{i,j}$.
 Then, it computes $T_{m_{i}} =  W \cdot \prod_{j=1}^{s} $ $g_{1}^{\beta^{j} \cdot (-\beta) \cdot \xi_{j} \cdot m_{i,j}} 
 =  W \cdot \prod_{j=1}^{s} g_{1}^{-\beta^{j+1} \cdot \xi_{j} \cdot m_{i,j}}$ and creates the MHT resulting from the file $m$ using $V$,
 such that if $(m_{i},\theta,V,W)$ exists in $L_{H'}$, then $W$ is used to compute $T_{m_{i}}$ and $V$ to construct the MHT; 
 otherwise, $\theta \in_{R} \mathbb{Z}_{p}$ is chosen at random, 
 $V=g_{1}^{\theta}$ and $W=h_{1}^{- \theta /\xi_{1}}$ are computed, 
 $(m_{i},\theta,V,W)$ is put in $L_{H'}$ and $W$ is used to compute $T_{m_{i}}$ and $V$ to construct the MHT. 
 It finally gets the corresponding root $rt$.
 It gives $T_{m}=(T_{m_{1}},\cdots,T_{m_{n}})$ and $\sigma_{rt} \gets \textsf{Sign}(sk_{\textsf{SS}},rt)$ to $\mathcal{A}$.  
  
\noindent $\diamond$
 \textit{Challenge.}
 $\mathcal{A}$ first gives to $\mathcal{B}$ two files $m_{0} =( m_{0,1} , \cdots , m_{0,n})$ and $m_{1} = ( m_{1,1} , \cdots, m_{1,n})$ of equal length and
 that have not been queried.
 $\mathcal{B}$ randomly selects a bit $b \in_{R} \{0,1\}$ and for $i \in [1,n]$, splits each block $m_{b,i}$ into $s$ sectors $m_{b,i,j}$.
 Then, it computes $T_{m_{b,i}} = W_{b} \cdot \prod_{j=1}^{s} $ $g_{1}^{-\beta^{j+1} \cdot \xi_{j} \cdot m_{b,i,j}}$ 
 and creates the MHT resulting from the file $m_{b}$ using $V_{b}$, such that
 if $(m_{b,i},\theta_{b},V_{b},W_{b})$ exists in $L_{H'}$, then $(V_{b},W_{b})$ is returned; 
 otherwise, $\theta_{b} \in_{R} \mathbb{Z}_{p}$ is chosen at random, 
 $V_{b}=g_{1}^{- \theta_{b}}$ and $W_{b}=h_{1}^{- \theta_{b} /\xi_{1}}$ are computed, 
  $(m_{b,i},\theta_{b},V_{b},W_{b})$ is put in $L_{H'}$ and $W_{b}$ is used to compute $T_{m_{b,i}}$ and $V_{b}$ to construct the MHT.
 It finally gets the corresponding root $rt_{b}$. 
 It gives the tag $T_{m_{b}}=(T_{m_{b,1}},\cdots,T_{m_{b,n}})$ 
 and the root's signature $\sigma_{rt_{b}} $ $\gets \textsf{Sign}(sk_{\textsf{SS}},rt_{b})$ to $\mathcal{A}$.
 
  Wlog, $\mathcal{A}$ generates a challenge on one block only. It chooses a subset $I=\{i^{*}\} \subseteq [1,n]$, 
 randomly chooses $v_{i^{*}} \in_{R} \mathbb{Z}_{p}$ and sets $chal=\{(i^{*},$ $v_{i^{*}})\}$.
 It forwards $chal$ as a challenge to $\mathcal{B}$.
 Upon receiving $chal$, $\mathcal{B}$ selects two ordered collections $F_{b}=\{m_{b,i^{*}}\}$ of blocks and 
 $\Sigma_{b}=\{T_{m_{b,i^{*}}}$ $\}$ which are the tags corresponding to the blocks in $F_{b}$ where
 $T_{m_{b,i^{*}}} =  W_{b} \cdot \prod_{j=1}^{s} g_{1}^{-\beta^{j+1} \cdot m_{b,i^{*},j} \cdot \xi_{j}}$.
It then randomly chooses $r_{j} \in_{R} \mathbb{Z}_{p}$, for $j \in [1,s]$ and computes 
 $R_{j}^{*} =h_{j}^{\beta^{2} \cdot r_{j}}=h_{j+2}^{\frac{\xi_{j}}{\xi_{j+2}} \cdot r_{j}}=g_{1}^{\beta^{j+2} \cdot \xi_{j} \cdot r_{j}}$, for $j \in [1,s-1]$
 and $R_{s}^{*} =Z^{\xi_{s} \cdot r_{s}}$. It implicitly fixes
$ b_{j}=\beta^{2} ( m_{b,i^{*},j} \cdot v_{i^{*}} + r_{j})$ for $j \in [1,s-1]$ 
by computing
 $B_{j}^{*} = h_{j}^{b_{j}}=$

\noindent $h_{j}^{\beta^{2} ( m_{b,i^{*},j} \cdot v_{i^{*}}+ r_{j})}= 
 h_{j+2}^{\frac{\xi_{j}}{\xi_{j+2}} (m_{b,i^{*},j} \cdot v_{i^{*}}+ r_{j})}= $ 

\noindent $g_{1}^{\beta^{j+2} \cdot \xi_{j}( m_{b,i^{*},j} \cdot v_{i^{*}} + r_{j})}$,
 and $B_{s}^{*} =Z^{\xi_{s} \cdot m_{b,i^{*},s} \cdot v_{i^{*}}} \cdot Z^{\xi_{s} \cdot r_{s}}$.
 It sets  as well \[ c^{*}  = T_{m_{b,i^{*}}}^{v_{i^{*}}} =  W_{b}^{v_{i^{*}}}  \cdot \prod_{j=1}^{s} g_{1}^{- \beta^{j+1} \cdot \xi_{j} \cdot m_{b,i^{*},j} \cdot v_{i^{*}}} .\]

 Finally, $\mathcal{B}$ returns $\nu^{*} = (R_{1}^{*} ,\cdots,R_{s}^{*} ,B_{1}^{*} ,\cdots,$ $B_{s}^{*} ,c^{*} )$ and $(V_{b},\Omega_{b,i^{*}})$, 
 where $\Omega_{b,i^{*}}$ are the AAI needed to create the MHT based on $V_{b}$.
Note that $\Omega_{b,i^{*}}$ is generated by calling successively $H'$. This means that, in the list $L_{H'}$, we can find tuples of the form
 $(z,\theta,V,W)$ such that the query $z$ can be either $m_{i}$ as we defined above (meaning that $m_{i}$ is appended to a leaf node) or $H'(y)$
 that is attached to an internal node.
 
 If $\mu = 0$ then $Z=g_{1}^{\beta^{s+2}}$. Thus, we have a valid random proof for $m_{b}$.
If $\mu = 1$, then $Z$ is random value in $\mathbb{G}_{1}$, and so the $R_{s}^{*}$ and $B_{s}^{*}$ are random elements in $\mathbb{G}_{1}$ 
from $\mathcal{A}$'s view and the proof of data possession contains no information about $m_{b}$.
  
\noindent $\diamond$
 \textit{Guess.} $\mathcal{A}$ returns a bit $b'$.
  If $b=b'$, $\mathcal{B}$ will output $\mu' = 0$ to indicate that it was given a $(s+1)$-DDHE tuple;
  otherwise it will output $\mu'=1$ to indicate that it was given a random tuple.

  \paragraph{Analysis.}
The tags and proof of data possession given to $\mathcal{A}$ are correctly distributed.
Indeed, when $\mu = 1$, $\mathcal{A}$ gains no information about $b$. Therefore, $Pr[b \neq b' | \mu =1] = 1/2$. 
Since $\mathcal{B}$ guesses $\mu'=1$ when $b \neq b'$, then $Pr[\mu = \mu' | \mu =1] = 1/2$.
If $\mu = 0$, then $\mathcal{A}$ sees an upload of $m_{b}$. 
$\mathcal{A}$'s advantage is thus negligible by definition (equal to a given $\epsilon$). 
Therefore, $Pr[b \neq b' | \mu =0] = 1/2 + \epsilon$. 
Since $\mathcal{B}$ guesses $\mu'=0$ when $b=b'$, we have $Pr[\mu = \mu' | \mu =0] = 1/2 + \epsilon$.
$T_{m_{b,i}}$ is equal to $(H'(m_{b,i}) \cdot \prod_{j=1}^{s} h_{j}^{m_{b,i,j}})^{-sk}$ 
where $sk$ is implicitly set as equal to $\beta$ and the values $h_{j}$'s are correctly distributed as in the real scheme.
Moreover, $sk_{\textsf{SS}}$ and $H'$ are kept secret from $\mathcal{A}$.
Note that $\mathcal{A}$ does not have access to $H'$ 
and the AAI given to $\mathcal{A}$ with the proof of data possession result from calls to $H'$.
In addition, $R_{j}^{*}$ and $B_{j}^{*}$ are statically indistinguishable with the actual outputs corresponding to $m_{0}$ or $m_{1}$.
Thus, the answers given to $\mathcal{A}$ are correctly distributed. The proof is completed.

\paragraph{N.B.} 
Such security level is reached for data privacy since $H'$ is kept secret by the server and the client and so, 
the adversarial TPA does not have access to it.

\subsection{Performance and Discussion with other existing works}

\noindent We first compare the MHT-based scheme with the original one presented in \cite{GSP15}.
The MHT-based construction seems less practical and efficient than the construction in \cite{GSP15}. 
Communication and computation burdens appear in order to obtain the desired security standards against the server and TPA.
The communication overheads increase between the client and server. 
The computation overheads for the client raise also, although the client is limited in resources.
The storage space of the server should be bigger, since it has to create and possibly stores MHTs for each client.
The TPA has to provide more computational resources for each client in order to ensure valid data integrity checks.
Nevertheless, experiments might show that the time gap between the algorithms in the scheme proposed in \cite{GSP15} and the ones in the MHT-based scheme is acceptable.
\\

The MHT is an Authenticated Data Structure (ADS) that allows the client and TPA to check 
that the server correctly stores and updates the data blocks. 

Erway et al. \cite{EKPT09} proposed the first DPDP scheme.
The verification of the data updates is based on a modified ADS, called Rank-based Authentication Skip List (RASL).
This provides authentication of the data block ranks, which ensures security in regards to data block dynamicity.
However, public verifiability is not reached. Note that such ADS with bottom-up leveling limits the insertion operations.
For instance, if the leaf nodes are at level $0$, any data insertion that creates a new level \textit{below} the level $0$ will bring necessary updates of all the
level hash values and the client might not be able to verify.

Wang et al. \cite{WWLRL09} first presented a DPDP with PV using MHT. However, security proofs and technical details lacked.
The authors revised the aforementioned paper \cite{WWLRL09} and proposed a more complete paper \cite{WWRLL11} 
that focuses on dynamic and publicly verifiable PDP systems based on BLS signatures.
To achieve the dynamicity property, they employed MHT. Nevertheless, because the check of the block ranks is not done, 
the server can delude the client by corrupting a challenged block as follows: it is able
to compute a valid proof with other non-corrupted blocks. Thereafter, in a subsequent work \cite{WWRL10}, 
Wang et al. suggested to add randomization to the above system \cite{WWRLL11}, in order to guarantee that the server cannot deduce
the contents of the data files from the proofs of data possession.

Liu et al. \cite{LRYZWC14} constructed a PDP protocol based on MHT with top-down leveling. Such protocol satisfies dynamicity and public verifiability.
They opted for such design to let leaf nodes be on different levels. Thus, the client and TPA have both to remember the total number of data blocks and check
the block ranks from two directions (leftmost to rightmost and vice versa) to ensure that the server does not delude the client with another node on behalf of a file
block during the data integrity checking process.

In this paper, the DPDP scheme with PV and DP is based on MHT with bottom-up leveling, such that data block ranks are authenticated.
Such tree-based construction guarantees secure dynamicity and public verifiability processes as well as preservation of data privacy, and remains practical in real environments.

\section{Conclusion}

\noindent We provided two solutions to solve the adversarial issues encountered in the DPDP scheme with PV and DP proposed in \cite{GSP15}.
These solutions manage to overcome replay attacks, replace attacks and attacks against data privacy
by embedding IHT or MHT into the construction in \cite{GSP15}.
We proved that the two new schemes are both secure against the server and data privacy-preserving against the TPA in the random oracle.

\section*{Acknowledgments}
\noindent This work was partially supported by the TREDISEC project (G.A. no 644412), funded by the European Union (EU) under the Information and Communication Technologies (ICT) theme of the Horizon 2020 (H2020) research and innovation programme.

\bibliographystyle{plain}
\bibliography{biblio}

\end{document}